\documentclass{aa} %
\usepackage{graphicx}
\usepackage{lscape}
\usepackage{longtable}
\usepackage{arydshln}
\usepackage{multirow}

\usepackage{epstopdf}
\usepackage{natbib}
\usepackage{txfonts}
\begin{document}
   \title{Spectroscopy of Very Low Mass Stars and Brown Dwarfs in the Lambda Orionis Star Forming Region.}

   \subtitle{II. On rotation, activity and other properties of spectroscopically confirmed members of Collinder 69. \thanks{Based on the ESO observing programs 080.C-0592 and  078.C-0124; and observing programs from Calar Alto, Keck, Subaru and Magellan.}}

   \author{A. Bayo
          \inst{1}
          \and
          D. Barrado
          \inst{2,3}
          \and
          N. Hu\'elamo
          \inst{3}
          \and
          M. Morales-Calder\'on          
          \inst{4}
          \and
          C. Melo
          \inst{1}
          \and
          J. Stauffer
          \inst{4}
          \and          
          B. Stelzer
          \inst{5}
                   }

   \institute{European Southern Observatory, Alonso de C\'ordova 3107, Vitacura, Santiago, Chile\\
              \email{abayo@eso.org}
         \and 
             Calar Alto Observatory, Centro Astron\'omico Hispano Alem\'an, C/Jes\'us Durb\'an Rem\'on, 2-2, 04004 Almer\'ia, Spain
         \and    
             Depto. Astrof\'isica, Centro de Astrobiolog\'ia (INTA-CSIC), P. O. Box 78, E-28691 Villanueva de la Ca\~nada, Spain
         \and
             Spitzer Science Center, California Institute of Technology, Pasadena, CA 91125
         \and
             INAF - Osservatorio Astronomico di Palermo, Piazza del Parlamento 1, 90134 Palermo, Italy}

   \date{}

 
  \abstract
   {Most observational studies so far point towards brown dwarfs sharing a similar formation mechanism as the one accepted for low mass stars. However, larger databases and more systematic studies are needed before strong conclusions can be reached.
   }
   {In this second paper of a series devoted to the study of the spectroscopic properties of the members of the Lambda Orionis Star Forming Region, we study accretion, activity and rotation for a wide set of spectroscopically confirmed members of the central star cluster Collinder 69 to draw analogies and/or differences between the brown dwarf and stellar populations of this cluster. Moreover, we present comparisons with other star forming regions of similar and different ages to address environmental effects on our conclusions.}
   {We study prominent photospheric lines to derive rotational velocities and emission lines to distinguish between accretion processes and chromospheric activity. In addition, we include information about disk presence and X-ray emission.
}
   {We report very large differences in the disk fractions of low mass stars and brown dwarfs ($\sim$58\%) when compared to higher mass stars  (26$^{+4}_{-3}$ \%) with 0.6 M$_{\odot}$ being the critical mass we find for this dichotomy. As a byproduct, we address the implications of the spatial distribution of disk and diskless members in the formation scenario of the cluster itself. We have used the H$\alpha$ emission to discriminate among accreting and non-accreting sources finding that  38$^{+8}_{-7}$ \% of sources harboring disks undergo active accretion and that his percentage stays similar in the substellar regime. For those sources we have estimated accretion rates. Finally, regarding rotational velocities, we find a high dispersion in v$\sin(i)$ which is even larger among the diskless population. 
   }	
   {}


   \maketitle
%

\section{Introduction}

This is the second paper of a series devoted to studying, from a spectroscopic point of view, the young population present in several associations belonging to the Lambda Orionis Star Forming Region (LOSFR). In this paper, we systematically analyze several properties of the confirmed members of the central cluster Collinder 69 (C69); including the presence of disks, accretion onto the central star/brown dwarf, rotation and activity.

In \cite{Bayo11} (from now on Paper I), we presented a very detailed and complete spectroscopic census of Collinder 69; the oldest ($\sim$ 5 -- 12 Myr) of the associations belonging to the LOSFR. In short, this star forming region is located at $\sim$400 pc \citep{Murdin77} representing the head of the Orion giant. Its center is dominated by the O8III multiple star $\lambda$ Orionis and comprises both recently formed stars from 0.2 M$_{\odot}$ to 24 M$_{\odot}$ and dark clouds actively forming stars.

The main goals of this paper are to study in detail the properties of the C69 confirmed members (e.g. rotation, activity, disk fractions and accretion rates), analyze the different populations within the cluster, and compare our results with other low mass star forming regions of similar and different ages. In fact, the evolutionary status of C69 seems to be specially suited to the study of disk evolution and accretion at the low end of the mass spectrum. According to our observations, in almost every bin in mass (for masses lower or equal to 0.6M$_{\odot}$, $\sim$3700K assuming a 5 Myr isochrone from \citealt{Baraffe98}), we find a diversity of disk harboring sources: from those with optically thick disks undergoing active accretion (onto the central star) to others that have completely lost their primordial circumstellar disks. 
Finally, and as a byproduct of this study, we try to put our findings in context of the current theory of how the LOSFR as a whole was formed (triggered by a supernova,
 see \citealt{DM01}, and more discussion on this scenario on Paper I).

This paper is organized as follows: In Section~\ref{sec:data} we provide a description of the data analyzed.
 In Sections~\ref{sec:rotvel} to ~\ref{sec:distrib} properties such as rotation velocity, activity levels, accretion processes, disks, variability and spatial distribution of the population of C69 are studied. And, finally, our summary conclusions are presented in Section~\ref{sec:conclusions}. 
 
 Unlike in Paper I, we have grouped the most interesting/puzzling objects into different categories (following the subsections of Sections~\ref{sec:rotvel} and~\ref{sec:acacc}) and discuss their peculiarities in Appendix A.


\section{Observations and data analysis}
\label{sec:data}

In this work we make use of an extensive spectral database that our group has been gathering during more than seven years and that is described in Paper I. Furthermore, we make use of the measurements and parameters derived for confirmed members published in \cite{DM99, DM01, Barrado07, Sacco08, Maxted08, MoralesPhD, BayoPhD, Barrado11} and \cite{Franciosini11}. 

\subsection{Spectroscopic database}

In total, we analyzed spectra obtained by us corresponding to $\sim$100 confirmed members, with several objects observed more than once, plus data corresponding to $\sim$70 objects that were studied in at least one of the papers cited before and for which we do not have our own observations.

In short, our own database comprises optical spectra of confirmed members of C69 with temperatures between $\sim$2800 and $\sim$4700 K. The spectra have resolutions in the range 600-11250, wavelength coverages from $\sim$5000~\AA~up to $\sim$10400~\AA~and have been obtained with different instruments and telescopes at Mauna Kea, Las Campanas, CAHA and VLT.

The data reduction of most of our campaigns was performed with IRAF\footnote{IRAF is distributed by the National Optical Astronomy Observatory, which is operated by the Association of Universities for Research in Astronomy, Inc. under contract to the National Science Foundation.} following the standard steps. For the analysis of the spectra, motivated by the large amount of data, we developed a tool that provides for any given set of lines in an automatic manner 
 (instrumental response corrected) full width half maximum (FWHM), full width at ten per cent of the flux (FW$_{10\%}$), and equivalent widths (EWs).
  The description of this routine and a study on the effect of the resolution on parameter determination is given in Appendix A of Paper I.

\subsection{Photometric and X-ray database}

In addition to the spectroscopic database, we have made use of the photometric dataset described in Paper I and analyzed in detail in \cite{MoralesPhD, BayoPhD} and Morales-Calderon et al 2012 in prep., and of the X-ray data presented in \cite{Barrado11} and \cite{Franciosini11}. The photometric database includes photometry from the optical to the mid-infrared (MIR). 

Taking a 5 Myr isochrone as a reference \citep{Baraffe98, Chabrier00,Allard03}, the completeness of the optical data is located at $\sim$20 M$_{\rm Jup}$ for the whole cluster and even down to $\sim$3~M$_{\rm Jup}$ for a field to the East of the center of the cluster. In the near infrared (NIR), the data is complete down to $\sim$10 M$_{\rm Jup}$ in almost the whole cluster and down to $\sim$30 M$_{\rm Jup}$ in the outskirts. For the MIR, our database is complete at 3.6 $\mu$m down to $\sim$40 M$_{\rm Jup}$.

Finally the X-ray data is complete down to ~$\sim$0.3M$_{\sun}$ with detections in \cite{Barrado11} for confirmed members as low as $\sim$0.1M$_{\sun}$.

\subsection{Fundamental parameters database}

We have adopted the results from \cite{MoralesPhD} and \cite{Barrado07} regarding the presence of circumstellar disks and substellar analogs. Objects with NIR and/or MIR slope above photospheric values are considered to harbor disks. This slope is analyzed as in \cite{Lada06} based on the $\alpha$ parameter.

The photometric database described before was used to feed VOSA \citep{Bayo08} and determine, via Spectral Energy Distribution (SED) fit, the effective temperatures and the bolometric luminosities for all confirmed members. These two parameters were used in Paper I, along with theoretical isochrones from the Lyon group \citep{Baraffe98, Chabrier00,Allard03}, to determine masses for the confirmed members. 

For the mass determination we assumed an age of 5Myr for C69 since that was the best fit to the produced HR diagram. We note, however, that in Paper I we also set an upper-limit for the age of C69 of 20Myr. We will address this point again in Section~\ref{subsec:diskfrac}.

\section{Rotational velocities}
\label{sec:rotvel}

For the sample of sources observed with the highest resolution (those observed with Magellan/MIKE, R$\sim$11250; 14 sources in total), we estimated their projected rotational velocity. We add to this sample the objects with $v\sin(i)$ determination in \cite{Sacco08} to study the rotation of $\sim$25 members of C69. 

\subsection{Technique}

\begin{center}
\begin{figure}
\includegraphics[width=9.0cm]{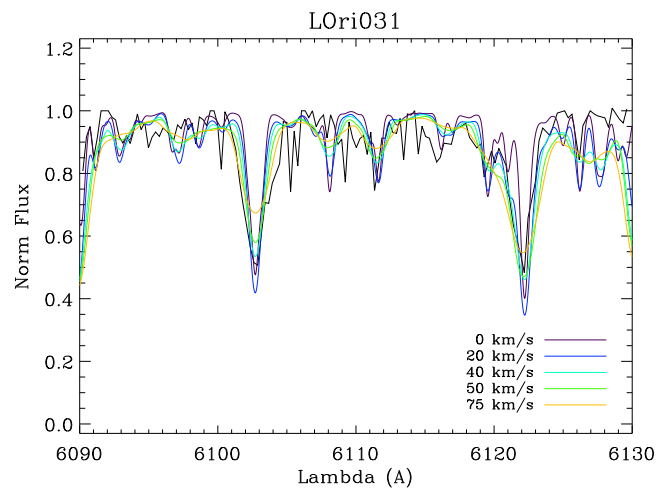}
\caption{Example of v$\sin(i)$ estimation for LOri031. In black we have plotted the science spectrum and on top of it with different colors the Kurucz model synthesized for 3750 K, log(g) of 4.0 and five different velocities. The synthetic spectra have been produced with the same resolution as the observed one.}
\label{vsini}
\end{figure}
\end{center}

We based our determination of the projected rotational velocity on the comparison of the observed spectra with Kurucz models \citep{Castelli97} synthesized for different rotational velocities. Since our estimation of $v\sin(i)$ is based on comparisons with models of a specific T$_{\rm eff}$, and there is a known dispersion between effective temperatures based on spectral types and those derived using models (T$_{\rm eff}$ scales, see Paper I), we followed a three step process:

\begin{enumerate}
\item First, we derived T$_{\rm eff}$ and log(g) by comparison with theoretical models. We built a grid of synthetic spectra using different collections: Kurucz \citep{Castelli97}, NextGen \citep{Hauschildt99} and Dusty \citep{Allard03} covering effective temperatures in the range 5000-2000 K, assuming solar metallicities, and a range of log(g) between 3.5 and 5.0. We lowered the resolution of both our spectra and the Kurucz models to R$\sim$200, which is the resolution provided in the public servers by the Lyon group, in order to perform a direct fitting process. 
Table~\ref{tab:paramMIKE} shows the best fitting set of parameters for each case.

\item Next, we prepared a second grid of high resolution Kurucz models (the ones we could generate with the same resolution as our Magellan/MIKE spectra), for different values of v$\sin(i)$ for direct comparison. 

For the comparison we chose the wavelength range from 6090~\AA~to 6130~\AA~because the signal-to-noise ratio (SNR) of our spectra was not very high in all cases, and, for the range of temperatures to be analyzed, two of the most prominent photospheric absorption lines (from Ca I) are located within this region. Besides, for the grid of high resolution models, we assume log(g) of 4.0 dex that is a suitable value (according to models) for these young cool confirmed members of C69.

\item Since our high resolution grid of models did not include models from the Lyon group, we only analyzed sources where Kurucz was the best choice in Step 1, or those for which the difference in reduced $\chi^2$ between Kurucz and NextGen was not significant (LOri050, LOri055, LOri057, LOri060 and LOri075).

We performed a model fit to the Ca I lines described in Step 2 (see Fig.~\ref{vsini}) in order to estimate the rotational velocities of our sources. Prior to this, we included the instrumental response measured on the MIKE lamp arc. The mean FWHM value measured on the lamp spectra was 0.4914~\AA, corresponding to a velocity of $\sim$20 km/s (the lower limit we can provide for several cases). 
\end{enumerate}

\subsection{Main results}

The estimated rotational velocities from this work are provided in Table~\ref{tab:paramMIKE}, and also in Table~\ref{tab:paramTOTAL} where we add those determined by \cite{Sacco08}. In general, the range of values determined in both studies agree very well. In particular, there are three sources for which both \citet{Sacco08} and us have estimated v$\sin(i)$. In two of these cases, LOri055 and LOri057, both estimations are in very good agreement, but in the case of LOri060 we provide a value $\sim$10 km/s higher than the one obtained by \citet{Sacco08}. For this source, our determination of the rotational velocity suffers from the fact that the S/N of the spectrum was the lowest of the sample and, for this reason, the wings of the Ca I doublet are not so ``clean" as in the other cases. Therefore, we take their upper-limit of $\sim$ 20 km/s as a more robust measurement.

\begin{figure}[htbp]
\begin{center}
\includegraphics[width=9.cm]{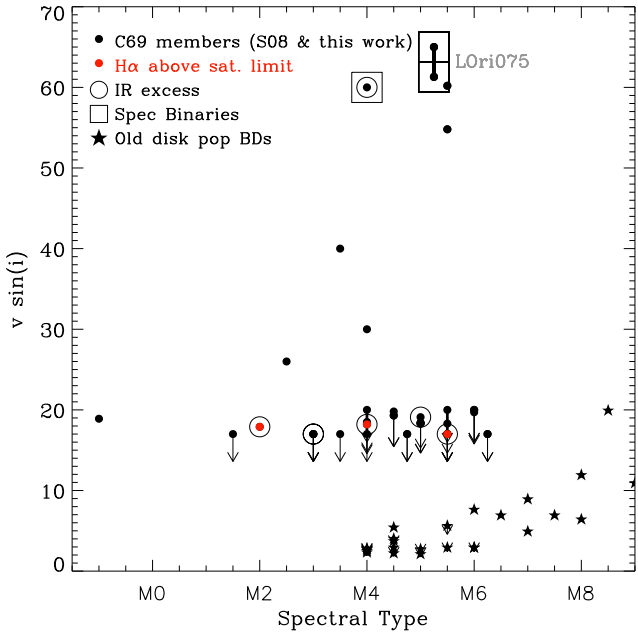}
\caption{Spectral Type vs v$\sin(i)$ for confirmed members of C69 from this work and \cite{Sacco08} (black dots). For comparison, we include the old disk population brown dwarfs from \cite{Mohanty03} with five pointed star symbols. From the C69 members we have highlighted those exhibiting peculiarities: large open squares surround spectroscopic binaries, large open circles represent objects with infrared excess (assumed to be a signpost of the presence of a disk) and red dots have been plotted on top of those objects classified as accretors according to the saturation criterion by \cite{Barrado03} (see text for details). Finally LOri075 is highlighted with a label and is discussed in Appendix~\ref{subsec:rotvel:ps}}
\label{fig:SpTvsini}
\end{center}
\end{figure}

In Fig.~\ref{fig:SpTvsini}, we display the projected rotational velocity as a function of spectral type for those objects where a value of v$\sin(i)$ (either calculated by us or from \citealt{Sacco08}) is available. Although a significant fraction of the measurements are upper limits, we can see how, as expected given the youth of the sample, the general trend for objects in C69 is to rotate faster than the old disk population of low-mass stars and brown dwarfs from \cite{Mohanty03} (starred symbols). 

Furthermore, although the dispersion of v$\sin(i)$ values among the members of C69 is large; if we consider the NIR and MIR slope as a proxy for disk presence, we see how this dispersion is larger among the diskless population than among disk-harboring sources. Although we are dealing with small number statistics, this could be the result of the 
disk locking effect \citep{Bouvier97}, that explains the dispersion in rotational velocities as a result of different decoupling timescales between the star and the disk.

\begin{table}
\caption{T$_{\rm eff}$, log(g) and rotational velocities derived by comparison with models for the sample of sources observed with Magellan/MIKE. We also display other relevant information about the sources such as accretion rate (derived in this work) or binarity.} 
\centering
\label{tab:paramMIKE}
\begin{tiny}
\begin{tabular}{@{\extracolsep{-7pt}}llllcccll}
\hline\hline
Source & Mdl$^{\mathrm{1}}$ & SpT & T$_{\rm eff}$&log(g)&v$\sin(i)$ &Acc$^2$&$\log(\dot{M}_{acc})$&Binary$^{3}$\\
              &             & (K)                   &            &  & (km/s)      &                                         & (M$_{\odot}$/yr)        &          \\
\hline
LOri017$^{\mathrm{*}}$ & Kur    & --       & 4250 & 4.0  & 70     &---      &                &\\
LOri031              	        & Kur    & M3.5 & 3750 & 4.0  & 40      & N &                &\\
LOri042              	        & Kur    & M4.0 & 3750 & 4.0  & 30      & N &                &\\
LOri050              	        & NG    & M4.0 & 3700 & 4.0  & 60      &Y&-10.878$\pm$0.05&M08, S08\\
LOri055              	        & NG   & M4.0 & 3700 & 4.0  & $<$20   & N &                &\\
LOri056              	        & NG   & M4.5 & 3700 & 4.0  & ---     & N &                &\\
LOri057              	        & NG   & M5.5 & 3700 & 4.0  & $<$20   & N &                &\\
LOri058              	        & NG   & M3.5 & 3500 & 4.0  & ---     & N &                &\\
LOri059              	        & NG   & M4.5 & 3500 & 4.0  &  ---     & N &                &\\
LOri060              	        & NG   & M4.0 & 3500 & 4.0  &  20-40      & N &                &\\
LOri063              	        & NG   & M4.0 & 3700 & 4.0  & ---     & Y &                &\\
LOri068              	        & NG   & M4.5 & 3700 & 4.0  & ---     & N &                &\\
LOri073              	        & NG   & M5.0 & 3700 & 4.0  & ---     & N &                &\\
LOri075              	        & NG   & M5.0 & 3400 & 4.0  & 65   & N &                &M08, B11\\
\hline\hline
\end{tabular}
\end{tiny}
\vspace{0.2cm}

\raggedright
\begin{footnotesize}
$^{\mathrm{*}}$There were problems on the extraction of several orders. We have used the T$_{\rm eff}$ corresponding to the SED fit provided in Paper I.\\
$^{1}$ Model collection corresponding to the best fitting model: NG = Nextgen; Kur = Kurucz\\
$^{2}$ Y indicates that the object fulfills the accretion criterion by \citet{Barrado03} (those marked with  N  do not).\\
$^{3}$ Binary system according to \citet[][M08]{Maxted08}, \citet[][S08]{Sacco08} or this work (B11).
\end{footnotesize}
\end{table}


\section{Activity and accretion}
\label{sec:acacc}

Activity, accretion, and mass loss processes can be studied through the analysis of emission lines in the spectrum of young stellar and substellar objects.
We can group these emission lines into two categories: (i) The forbidden emission lines of [OI] ($\lambda$5577, 6300, 6364 \AA), [SII] ($\lambda$6717, 6731 \AA) and [NII] ($\lambda$6548, 6581 \AA) that have been attributed in the literature to low density regions such as winds, and are therefore a tracer of the mass loss process (see \citealt{Shu94,Hartmann94} and \citealt{Hartmann99}). (ii) The permitted emission lines of He I ($\lambda$6678~\AA), H$\alpha$ ($\lambda$6563~\AA) and the CaII triplet ($\lambda$8498, 8542, 8662~\AA) which are characteristic of classical T Tauri stars and their substellar analogs and trace accretion processes, although they are also known to be signposts of chromospheric activity \citep{Martin01, Natta01, Natta02, Testi02, Mohanty03b, Jayawardhana02a,Jayawardhana02b, Jayawardhana03a, Jayawardhana03b, Barrado02, Barrado03b,White03}. 

Given the evolutionary status of the members of C69 and the environment where they are located, most of the analyzed spectra show a rich variety of emission lines. We summarize the results from our analysis in the following subsections.

\subsection{Spectroscopic emission lines among the C69 sample}
\label{subsec:el}

For our sample, the forbidden emission lines (given their narrow nature) were only detected in some of the higher resolution spectra (see tables~\ref{tab:emissionlinesMIKELRIS} and~\ref{tab:emissionlinesLRISFLAMES}), but since most of these detections are quite marginal we cannot be certain whether the signal comes from the source itself or from the nebular environment.

The Ca II triplet was covered in very few of our runs and only detected in one object (C69-IRAC-005) that is analyzed in detail in Appendix~\ref{subsubsec:Haacc:ps}. 

On the other hand, we have measurement/s of H$\alpha$ for 156 sources out of the 172 spectroscopically confirmed members. The vast majority of these sources(140 objects, $\sim$90\% of the sample) show the line in emission. The EW of this line is used to distinguish between chromospheric or accretion origin of the emission in Section~\ref{subsec:Haacc}.

Regarding He I; we have detected this line in 24 objects, all in emission, and we provide the EWs in tables~\ref{tab:emissionlinesMIKELRIS} and~\ref{tab:emissionlinesLRISFLAMES}. All these sources present also intense H$\alpha$ emission (with a minimum EW of $\sim$5\AA) and six of them are classified as accretors in Section~\ref{subsec:Haacc}. Regarding the mass range where He I is detected; only six of the sources are in the BD domain and the rest have masses above the hydrogen burning limit. 

Further comparison of the He I and H$\alpha$ EWs of the 24 sources with He I detection shows a clear correlation among them; the stronger the H$\alpha$ emission, the stronger the He I one; but our sample is not statistically significant to derive any quantitative relationship between them.

The fact that we detect He I mainly in stars and not in brown dwarfs is not surprising since the typical structure of the He I line is much narrower than that of H$\alpha$, for example, and therefore this emission can be hidden if the spectrum does not have high SNR and resolution.

\begin{figure}[htbp]
\begin{center}
\includegraphics[width=9.cm]{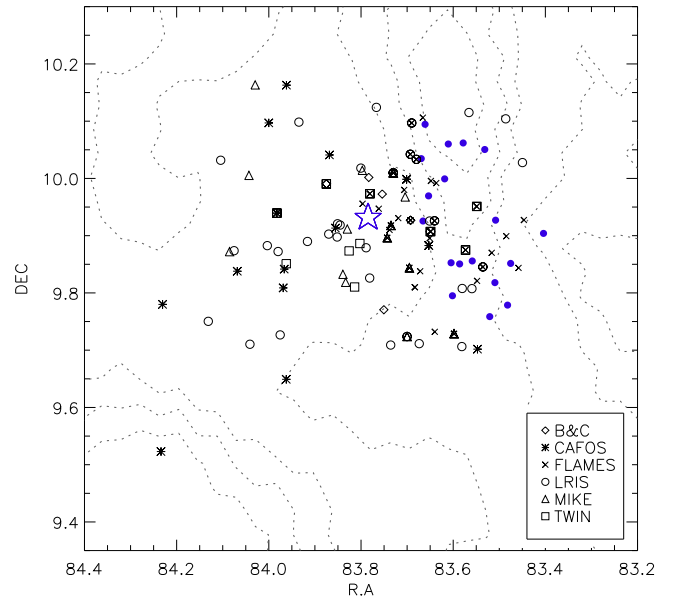}
\caption{Contours corresponding to an H$\alpha$ image of C69 (extracted from the H$\alpha$ six arcmin resolution all sky survey compiled by \citealt{Finkbeiner03}). We have plotted the location of the C69 members using different symbols (all of them in black, see legend of the figure) according to the instrument with which the individual spectrum was obtained. As can be seen a significant fraction of the sources have been observed more than once by us. The locations of sky fibers in the VLT/FLAMES campaign are displayed as blue filled dots and the position of Lambda Orionis (roughly marking the center of the cluster) is highlighted with a large blue five-pointed star.
}
\label{fig:HalphaFLAMES}
\end{center}
\end{figure}

In addition to this technical caveat to detect He I, the extended nebular emission of the cluster has to be taken into account when dealing with fiber spectrographs. In particular,  in two cases from the VLT/FLAMES spectra, the helium line could be contaminated by nebular emission (see in Fig.~\ref{fig:HalphaFLAMES} the distribution of the sky fibers and in Section~\ref{subsec:Haacc} the quantification of this contamination for the H$\alpha$ line). 

\begin{table*}[]
\tiny
\begin{center}
\caption{Equivalent width (negative for emission, positive for absorption) of the main lines observed in the spectra obtained with Magellan/MIKE and Keck/LRIS in its low and high resolution modes (the three different setups are separated by a double line). 
We only show objects where at least one of the forbidden lines, and/or H$\beta$, and/or HeI have been measured.
The measurements of H$\alpha$ and the Ca II triplet are shown in Table~\ref{tab:paramTOTAL} (with the opposite convention: positive for emission).
The symbol ``--'' indicates that the line has not been detected while a white space is indicative of the line falling outside of the wavelength coverage of the spectra. Spectral types (SpT) from Paper I (averaged when more than one spectral type was given).
}
\label{tab:emissionlinesMIKELRIS}
    \begin{tabular}{@{\extracolsep{-4pt}}lcccccccccc}
\hline\hline
Object        & SpT  &   H$\beta$     &  [OI]          &  [OI]          &   [OI]        &  [NII]        &  [NII]         & HeI                 &  [SII]         & [SII]           \\
              &      &   4861         &  5577          &  6300          &   6364        &  6548         &  6581          & 6678                &  6717          & 6731            \\
              &      &   EW(\AA) eEW  &  EW(\AA) eEW   &  EW(\AA)  eEW  &   EW(\AA) eEW &  EW(\AA)  eEW &  EW(\AA)  eEW  & EW(\AA)  eEW        &  EW(\AA)  eEW  & EW(\AA)   eEW   \\
\hline        
LOri050       & M4.0 & -9.41$\pm$1.64 & -0.53$\pm$0.04 & -0.58$\pm$0.03 &            -- &            -- &             -- & -0.44$\pm$0.07      &             -- &               -- \\ 
LOri055       & M4.0 & -5.11$\pm$0.30 &             -- &             -- &            -- &            -- &             -- &                  -- &             -- &               -- \\ 
LOri056       & M4.5 & -5.16$\pm$0.78 &             -- &             -- &            -- &            -- &             -- &                  -- &             -- &               -- \\ 
LOri060       & M4.0 & -4.13$\pm$0.47 &             -- & -1.14$\pm$0.42 & 0.91$\pm$0.10 &            -- &             -- &                  -- &             -- &               -- \\ 
LOri063       & M4.0 &             -- &             -- &             -- &            -- & 0.73$\pm$0.05 & -2.51$\pm$0.39 &                  -- & -0.93$\pm$0.22 & -0.78$\pm$0.07   \\ 
LOri068       & M4.75& -5.49$\pm$0.62 &             -- &             -- &            -- &            -- &             -- &                  -- &             -- &               -- \\ 
LOri075       & M5.25&             -- & -5.47$\pm$1.53 & -1.76$\pm$0.50 &            -- &            -- &             -- &                  -- &             -- &               -- \\ 
\hline
\hline
LOri098       & M5.0 &                &                &             -- &            -- &            -- &             -- &  -2.24$\pm$0.58     &       --       &              --  \\ 
LOri107       & M6.0 &                &                &             -- &-0.68$\pm$0.13 &            -- &             -- &              --     &       --       &              --  \\ 
LOri114       & M6.0 &                &                & -2.39$\pm$0.86 &            -- &            -- &             -- &              --     &       --       &              --  \\ 
LOri115       & M5.0 &                &                & -0.95$\pm$0.03 &            -- &            -- &             -- &              --     &       --       &              --  \\ 
LOri139       & M5.75&                &                & -2.76$\pm$0.28 &            -- &            -- &             -- &  -0.88$\pm$0.18     &       --       & -1.13$\pm$0.30   \\ 
LOri140       & M7.0 &                &                &-11.82$\pm$3.64 &            -- &            -- &             -- &  -4.24$\pm$0.17     &       --       &             --   \\ 
LOri155       & M8.0 &                &                &             -- &            -- &-4.28$\pm$0.82 & -4.42$\pm$0.55 &  -2.81$\pm$0.64     &       --       &             --   \\ 
\hline\hline     
LOri068       & M4.75&                &                &                &               &            -- &            --  &  -1.42$\pm$0.24     & -0.76$\pm$0.38 & -0.41$\pm$0.18   \\ 
LOri071       & M5.0 &                &                &                &               &-0.49$\pm$0.11 & -1.13$\pm$0.59 &                 --  &            --  & -0.07$\pm$0.01   \\ 
LOri077       & M5.0 &                &                &                &               &-0.72$\pm$0.25 &            --  &                 --  & -1.14$\pm$0.52 &              --  \\ 
LOri089       & M5.0 &                &                &                &               &-0.42$\pm$0.11 &            --  &                 --  & -0.82$\pm$0.38 &              --  \\ 
LOri091       & M4.75&                &                &                &               &            -- & -0.11$\pm$0.01 &  -0.95$\pm$0.31     & -0.95$\pm$0.69 & -0.34$\pm$0.10   \\ 
LOri094       & M5.5 &                &                &                &               &            -- &            --  &  -1.87$\pm$0.39     &            --  & -1.21$\pm$0.82   \\ 
LOri099       & M5.5 &                &                &                &               &-0.58$\pm$0.12 &            --  &                 --  &            --  &              --  \\ 
LOri106       & M5.5 &                &                &                &               &-0.58$\pm$0.19 &            --  &  -1.01$\pm$0.25     & -0.19$\pm$0.02 & -0.26$\pm$0.04   \\ 
LOri109       & M5.75&                &                &                &               &            -- &            --  &  -1.14$\pm$0.22     &            --  & -0.15$\pm$0.04   \\ 
LOri113       & M5.5 &                &                &                &               &            -- & -0.17$\pm$0.05 &  -1.10$\pm$0.14     & -1.73$\pm$0.42 &              --  \\ 
LOri119       & M5.5 &                &                &                &               &-4.16$\pm$1.03 &            --  &  -0.58$\pm$0.19     & -0.25$\pm$0.08 &              --  \\ 
LOri129       & M6.0 &                &                &                &               &            -- &            --  &  -0.68$\pm$0.15     &            --  &              --  \\ 
\hline\hline
\end{tabular}
\end{center}
\end{table*}

\begin{table}[]
\tiny
\begin{center}
\caption{Equivalent width of the main lines observed in the spectra obtained with VLT/FLAMES. 
We follow the same conventions as in Table~\ref{tab:emissionlinesMIKELRIS}.
}
\label{tab:emissionlinesLRISFLAMES}
    \begin{tabular}{@{\extracolsep{-8pt}}llccccc}
\hline\hline
Object        & SpT  &  [NII]          & HeI             &   [SII]        & [SII]          \\
              &      &  6581           & 6678            &   6717         & 6731           \\
              &      &  EW(\AA)   eEW  & EW(\AA)  eEW    &   EW(\AA)  eEW & EW(\AA)   eEW  \\
\hline          
C69-IRAC-006  & M3.5 & -0.69$\pm$0.06  &            --   & -0.39$\pm$0.13 & -0.31$\pm$0.05 \\ 
C69-IRAC-007  & M2.5 &             --  & -0.48$\pm$0.15  &             -- & -0.20$\pm$0.04 \\ 
LOri038       & M3.0 &             --  & -0.45$\pm$0.13  &             -- &             -- \\ 
LOri043       & M4.0 &             --  &            --   & -0.26$\pm$0.07 &             -- \\ 
LOri045       & M3.0 & -0.40$\pm$0.04  & -0.34$\pm$0.07  &             -- &             -- \\ 
LOri050       & M4.0 & -0.45$\pm$0.06  & -0.63$\pm$0.24  &             -- &             -- \\ 
LOri053       & M5.0 &             --  &            --   & -0.25$\pm$0.03 &             -- \\ 
LOri059       & M4.5 & -0.62$\pm$0.13  & -0.31$\pm$0.05  & -0.68$\pm$0.13 & -0.36$\pm$0.08 \\ 
LOri069       & M5.5 & -0.76$\pm$0.13  &            --   & -0.46$\pm$0.11 & -0.50$\pm$0.10 \\ 
LOri073       & M5.25& -0.37$\pm$0.09  & -0.22$\pm$0.03  &             -- &             -- \\ 
LOri075       & M5.25& -0.73$\pm$0.17  &            --   &             -- &             -- \\ 
LOri079       & M6.25& -1.43$\pm$0.07  & -0.40$\pm$0.06  &             -- & -0.67$\pm$0.04 \\ 
LOri086       & M5.0 & -1.12$\pm$0.13  & -0.34$\pm$0.05  &             -- &             -- \\ 
LOri087       & M5.0 & -0.58$\pm$0.13  & -0.51$\pm$0.03  & -0.90$\pm$0.05 & -0.41$\pm$0.08 \\ 
LOri088       & M5.5 & -1.64$\pm$0.13  & -0.51$\pm$0.09  & -1.59$\pm$0.37 & -0.72$\pm$0.18 \\ 
LOri091       & M4.75& -0.44$\pm$0.06  &            --   & -0.81$\pm$0.36 &             -- \\ 
LOri093       & M5.5 & -1.00$\pm$0.06  & -0.24$\pm$0.03  &             -- &             -- \\ 
LOri094       & M5.5 & -1.71$\pm$0.21  & -0.41$\pm$0.07  &             -- & -0.98$\pm$0.19 \\ 
LOri105       & M6.0 & -2.68$\pm$0.52  &            --   &             -- & -0.72$\pm$0.19 \\ 
LOri109       & M5.75& -2.64$\pm$0.44  &            --   & -1.42$\pm$0.17 & -1.04$\pm$0.13 \\ 
LOri112       & M6.5 & -1.76$\pm$0.20  &            --   & -2.07$\pm$0.25 &             -- \\ 
LOri115       & M5.0 &             --  & -0.83$\pm$0.08  & -0.98$\pm$0.29 &             -- \\ 
LOri120       & M5.5 & -3.31$\pm$0.17  &            --   & -4.73$\pm$0.55 &  1.34$\pm$0.12 \\ 
\hline\hline
\end{tabular}
\end{center}
\end{table}

\subsection{Variability connected to activity}
\label{subsec:var}

Some hints on the chromospheric activity among the late-type population of C69 have already been studied in Paper I through the analysis of the alkali lines. 
Another clear signpost of activity in young stellar and substellar objects is variability. This variability has been observed not only in the continuum but also in a variety of lines, and its dependence with the spectral type has also been addressed in the literature (examples can be found in \citealt{Soderblom93, Stauffer97, Barrado01}). In Paper I, we reported a large fraction ($\gtrsim$35\%) of the members of C69 showing such variability in alkali lines (lithium and sodium).

Most of the sources showing variability in the alkali lines also show variability in the H$\alpha$ emission, and in Paper I we already demonstrated that the differences we see in, for example the EW of the lines, are larger than those we would expect just because we are comparing observations with different spectral resolutions. 

\begin{figure}
\centering
\includegraphics[width=8.0cm]{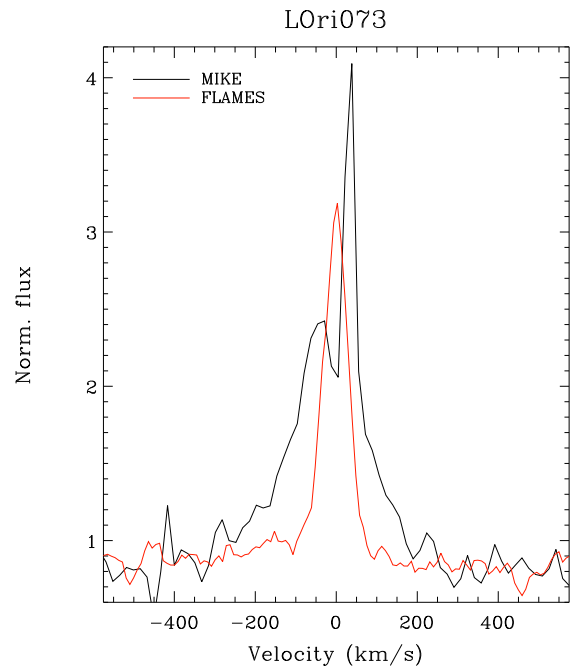}
\caption{H$\alpha$ profile of LOri073 observe at high resolution in two different epochs: with VLT/FLAMES (in red, R$\sim$8600) in Jan 2008 and with Magellan/MIKE (in black, R$\sim$11250) in Dec 2002.}
\label{fig:varHa}
\end{figure}

On the other hand, we find a couple of objects exhibiting significant variability in H$\alpha$ and no variability in the alkali lines (see Table~\ref{Tab:variability}, the top set of sources). The most extreme example is LOri073; this object not only shows variations in the measured EW of H$\alpha$, but also in the profile of the line. As can be seen in Fig.~\ref{fig:varHa}, while we observed a single peak line profile with VLT/FLAMES, a double peak structure was observed several years before with Magellan/MIKE. This more complex profile is very similar, although narrower, to that presented in \cite{Fernandez04} for the weak-line T Tauri star V410 Tau: we can see the narrow emission peak slightly red-shifted from the rest velocity, as well as the shallower component and the blue-shifted absorption that could suggest the presence of a wind.

Finally, another three particularly interesting variable objects are discussed in Appendix~\ref{subsec:var:ps}. They present the peculiarity that according to one of the H$\alpha$ measurements, the emission is too large to have a purely chromospheric activity origin (see Table~\ref{Tab:variability}, second sub-set of objects), but for all the other measurements this is not the case.

\begin{table*}[]
\tiny
\begin{center}
\caption{Parameters measured and calculated for those sources showing variations in H$\alpha$ and no significant variability in alkali lines; or variations in H$\alpha$ that causes them to oscillate between accreting and non-accreting (see text for details). Positive equivalent widths for H$\alpha$ denote emission while for the alkali lines it means absorption. We compile also the values from \cite{Sacco08} regarding H$\alpha$ and Li I.}
\label{Tab:variability}
    \begin{tabular}{lcccccccccc}
\hline\hline
Name            & EW(H$\alpha$)   & EW(H$\alpha$) S08  &   Ins. Res         & Mass(M$_{\odot}$) &Class& Disk &Li I & Li I S08         &  KI              & NaI\\
\hline                                                                                
\multirow{2}{*}{LOri073}         & -10.91$\pm$1.04 &  &  MIKE R$\sim$11250 & \multirow{2}{*}{0.34} & \multirow{2}{*}{III}  & \multirow{2}{*}{Diskless}  &0.62$\pm$0.03 &                                      &                  & \multirow{2}{*}{2.5$\pm$0.02$^{**}$} \\
                                 & -4.54 $\pm$0.19 &  				   &  FLAMES R$\sim$8600  &                       &                       &                            &    0.66$\pm$0.05          &              &                  &  \\
\hdashline                                                                                                          	                  
\multirow{2}{*}{DM048}         & -3.36$\pm$0.11 &   & CAFOS R$\sim$600 & \multirow{2}{*}{0.25} & \multirow{2}{*}{II}  & \multirow{2}{*}{Thick}  & &   &  0.29$\pm$0.07                &  2.57$\pm$0.43\\
                                                    & -7.72 $\pm$0.19  &   &  TWIN R$\sim$1100  &                       &                       &                            &              &              &   0.49$\pm$0.21               & 2.36$\pm$0.18 \\
\hline\hline
\multirow{2}{*}{LOri068}         &  -6.82$\pm$0.52 &\multirow{2}{*}{-7.9$\pm$1.26}   &  MIKE R$\sim$11250 & \multirow{2}{*}{0.35} & \multirow{2}{*}{III}  & \multirow{2}{*}{Diskless}  &0.69$\pm$0.09 &\multirow{2}{*}{0.698$\pm$0.025} &                  &  \\
                                 & -14.39$\pm$0.74 &  			 	   &  LRIS R$\sim$2650  &                        &                       &                            &0.53$\pm$0.05 &              &                  &  \\
\hdashline                                                                                                          	                  
\multirow{2}{*}{LOri075}         & -12.81$\pm$0.86 &\multirow{2}{*}{-10.71$\pm$0.98} &  MIKE R$\sim$11250 & \multirow{2}{*}{0.25} & \multirow{2}{*}{III}  & \multirow{2}{*}{Diskless}  &0.94$\pm$0.05 &\multirow{2}{*}{0.455$\pm$0.044} &                  &  \\
                                 & -10.20$\pm$0.89 &  				   &  LRIS R$\sim$950  &                       &                       &                            &              &              &                  &  2.82$\pm$0.38\\
\hdashline                                                                                                          	                  
\multirow{2}{*}{LOri080}         & -21.3$\pm$1.00 &\multirow{2}{*}{-14.47$\pm$1.25}  &     BC R$\sim$2600 & \multirow{2}{*}{0.15} & \multirow{2}{*}{III}  & \multirow{2}{*}{Diskless}  &              &\multirow{2}{*}{0.523$\pm$0.020} &  1.66$\pm$0.89   &  \\
                                 & -22.82$\pm$0.62 &  				    &   TWIN R$\sim$1100 &                       &                       &                            &              &              &  1.77$\pm$0.33   &  2.34$\pm$0.46\\
\hdashline                                                                                                          	                  
\multirow{3}{*}{LOri091$^{*}$}         & -11.93$\pm$1.12 &  				    &   LRIS R$\sim$2650 & \multirow{3}{*}{0.13} & \multirow{3}{*}{III}  & \multirow{3}{*}{Diskless}  &0.23$\pm$0.07 &              &  2.17$\pm$0.46   &  \\
                                 & -23.2$\pm$12.3  &  				    &   TWIN R$\sim$1100 &                       &                       &                            &              &              &                  &  2.05$\pm$0.70\\
                                 & -13.47$\pm$1.11 &  				    & FLAMES R$\sim$8600 &                       &                       &                            &0.76$\pm$0.06 &              &                  &  \\
\hdashline                                                                                                          	                  
\multirow{2}{*}{LOri109}         & -19.62$\pm$0.52 &  				    & FLAMES R$\sim$8600 & \multirow{2}{*}{0.16} & \multirow{2}{*}{III}  & \multirow{2}{*}{Diskless}  &0.57$\pm$0.11 &              &                  &  \\
                                 &  -8.70$\pm$0.45 &  				    &  LRIS R$\sim$2650  &                       &                       &                            &0.50$\pm$0.15 &              &  2.62$\pm$0.57   &  \\ 
\hline\hline
\end{tabular}
\end{center}
\raggedright
\begin{tiny}
$^{\mathrm{*}}$The signal to noise ratio of the TWIN spectrum of LOri091 is very low. This is clear in the very large error on the determination of the EW. Therefore this object is not considered variable.\\
$^{\mathrm{**}}$Value from \cite{Maxted08}.\\
\end{tiny}
\end{table*}

\subsection{H$\alpha$ emission as a proxy for accretion}
\label{subsec:Haacc}


We note once again that the H$\alpha$ emission can have different origins, and although it is commonly used as a proxy for active accretion, some considerations have to be taken into account depending on the nature of the source and its surroundings.

\begin{figure}[htbp]
\begin{center}
\includegraphics[width=9.cm]{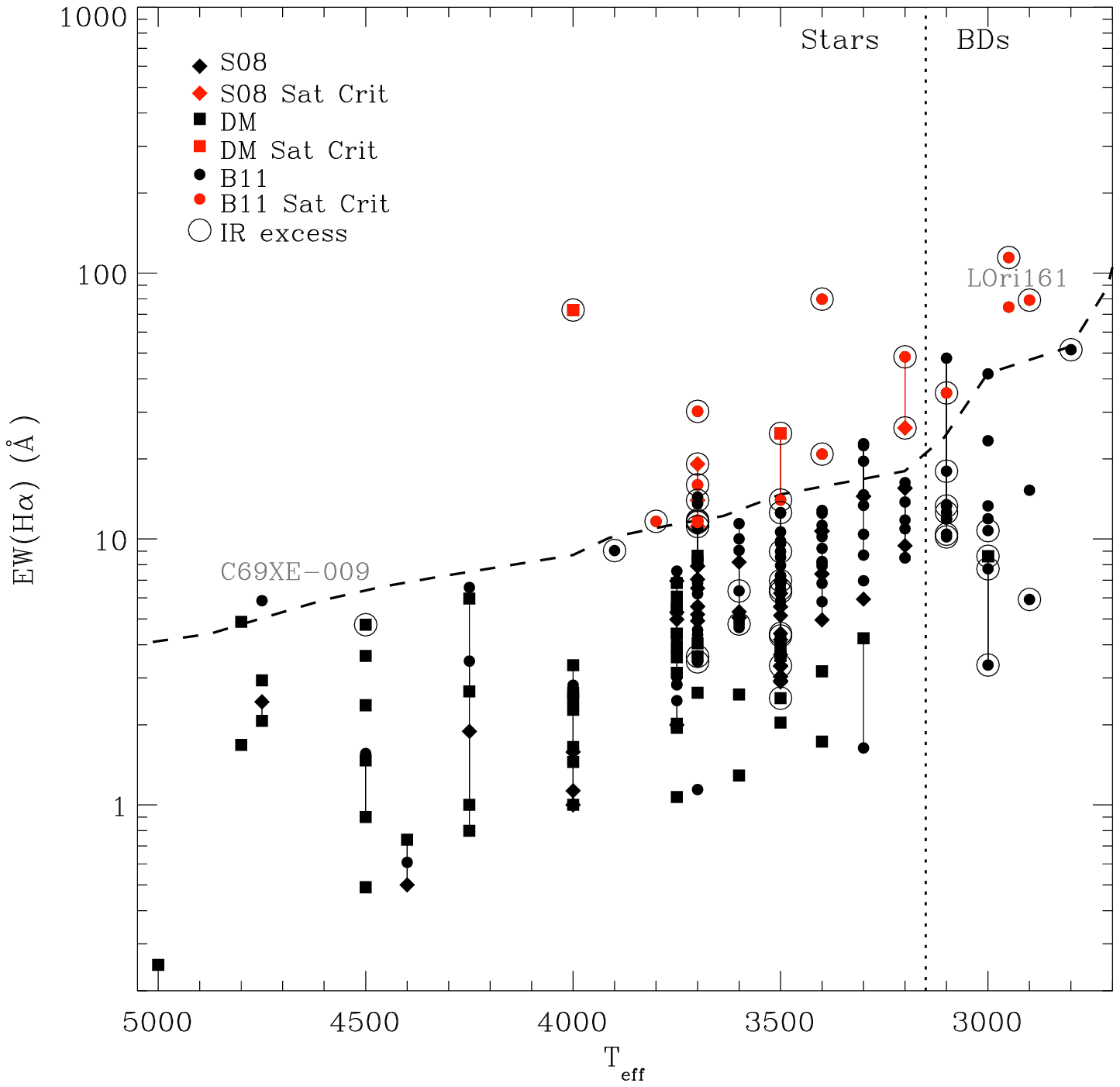}
\caption{H$\alpha$ equivalent width versus the effective temperature for confirmed members of C69. We display data from \citet{DM99, DM01} with filled squares, and from \cite{Sacco08} with filled diamonds. Our data are displayed with solid circles. In every sample, red symbols are used for those sources classified as accretors. Overlapping large circles highlight sources exhibiting excess in the Spitzer/IRAC photometry. For some sources we had more than one epoch of data (either ours or from \citealt{DM99,DM01,Sacco08}). These different measurements for individual objects appear joint with a solid line. The short-dashed line corresponds to the saturation criterion defined by \citet{Barrado03}. A vertical dotted line highlights the substellar frontier for an estimated age of 5 Myr according to the isochrones from the Lyon group. Particular sources discussed in Appendix~\ref{subsubsec:Haacc:ps} are highlighted with grey labels.}
\label{fig:HaSpT}
\end{center}
\end{figure}

First of all, and since cool objects are known to be very active, the H$\alpha$ emission can be chromospheric and not related to accretion processes.
 Some limits to this chromospheric contribution have been suggested in the literature: \cite{White03} proposed EWs of 10\AA~and 20\AA~for spectral types K3-M2 and later than M2, respectively. On the other hand, \citet{Barrado03} proposed a more spectral dependent relationship mainly focused on late K, M and L dwarfs (and therefore more suitable for our study). This empirical criterion is based on the saturation limit of chromospheric activity ($[L({\rm H}\alpha)/L_{\rm bol}$] = -3.3) and it is shown in Fig~\ref{fig:HaSpT} as the boundary to discriminate between accreting and non-accreting stars and brown dwarfs.

On the other hand, when trying to extract accurate measurements from the H$\alpha$ emission to disentangle between activity and accretion, we are also forced to take into account the environment where the sources are located. As mentioned before, clusters like C69, possess a non-negligible nebular component (see Fig.~\ref{fig:HalphaFLAMES}), and therefore one has to make sure that the nebular component is subtracted properly from the spectrum of each science target. This is not an issue, for instance, when dealing with long-slit spectroscopy (like most of our campaigns) but can induce some bias when observing with fiber spectrographs if the ``sky fibers" are not placed carefully.

In Fig.~\ref{fig:HalphaFLAMES} we show the contours corresponding to an H$\alpha$ image (from \citealt{Finkbeiner03}, with low spatial resolution $\sim$6') of the LOSFR, where we have highlighted with crosses the sources observed with VLT/FLAMES and with blue filled circles the locations of the sky fibers used to correct our measurements. At first sight, some structure can be inferred in the nebular emission surrounding our science targets. To better characterize this effect, we studied the variations of the H$\alpha$ nebular emission with the sky fibers: we measured a mean full width at 10\% of $\sim$41 km/s with a standard deviation of $\sim$3 km/s. We computed a median sky fiber and corrected the science spectra with that median; therefore the dispersion measured in those fibers translated into an added $\sim$7\% uncertainty in our measurements.

The original accretion criterion provided in \citet{Barrado03} shows the limiting EW(H$\alpha$) as a function of the spectral type. Since for most of the sources from \cite{DM99,DM01} and a fraction of those from \citealt{Sacco08} we do not have spectral type determination, we have used the temperature scale derived in Paper I to translate the original criterion into a T$_{\rm eff}$ vs. EW(H$\alpha$) relationship. 

As mentioned in Section~\ref{subsec:el}, from the total census of 172 spectroscopically confirmed members of C69, there are 16 sources for which we do not have a measurement of the equivalent width of H$\alpha$ and therefore we cannot apply the criterion.
Those 16 sources present very dispersed properties such as effective temperature, disk presence, etc; and therefore they should not produce any bias in the statistics derived for the whole cluster.

In Fig~\ref{fig:HaSpT} we show the accretion criterion applied to the 156 confirmed members with measurements of the EW(H$\alpha$). In order to be consistent; for objects having a spectral type determination, we have also translated them into effective temperatures using the temperature scale from Paper I (as we did for the criterion itself). On the other hand, for objects without spectral type determination, we have assumed the T$_{\rm eff}$ derived from the SED fit performed with VOSA (\citealt{Bayo08}, 2012, submitted). 
Applying the saturation criterion, we classified 9$^{+3}_{-2}$\% of the members as accretors (14 red dots in the figure; see column ``Acc" from Table~\ref{tab:paramTOTAL}). 

To relate this percentage to the disk presence, in  Fig~\ref{fig:HaSpT}, we have also included the information regarding the infrared class (as a proxy for the presence of disk, from \citealt{Barrado07} and Morales-Calder\'on et al. 2012, in prep.) with larger circumferences surrounding Class II sources and/or those with a MIR slope compatible with thick, thin or transition disks. 
If we only consider the 37 members showing signposts of harboring disks and with available measurements of EW(H$\alpha$), we estimate that 38$^{+8}_{-7}$ \% show active accretion according to the saturation criterion.

To focus now on the substellar population of C69; if we consider substellar those sources with estimated masses lower than 0.1M$_{\odot}$ to take into account uncertainties in the mass determination, we find eight brown dwarfs harboring disks and three of them to be accreting according to the criterion. 
This leaves us with a fraction of 37.5$^{+18}_{-13}$  \% of accretors among the disk-beating BDs, very similar to that of the stellar population and comparable with the substellar one provided in \cite{Scholz07} for the similar age cluster Upper Sco (31\%, 4 our of 13 objects).

In Table~\ref{tab:paramTOTAL}, we present the measured EWs 
 of the H$\alpha$ line of the sources in the sample. For those classified as accretors, 
 we have estimated the mass accretion rate using the measured FW$_{10\%}$(H$\alpha$) and the following relationship derived by \citet{Natta04} \footnote{The FW$_{10\%}$(H$\alpha$) of the sources are typically above the threshold of 200 km/s determined by \citet{Natta04} and both quantities, accretion rates and FW$_{10\%}$(H$\alpha$), are provided in Table~\ref{tab:paramTOTAL}}:

\begin{equation}
\log(\dot{M}_{acc}) = -12.89(\pm0.3)+9.7(\pm0.7)\times 10^{-3} {\rm FW}_{10\%}(H\alpha)
\end{equation}

\begin{figure}
\centering
\includegraphics[width=9.0cm]{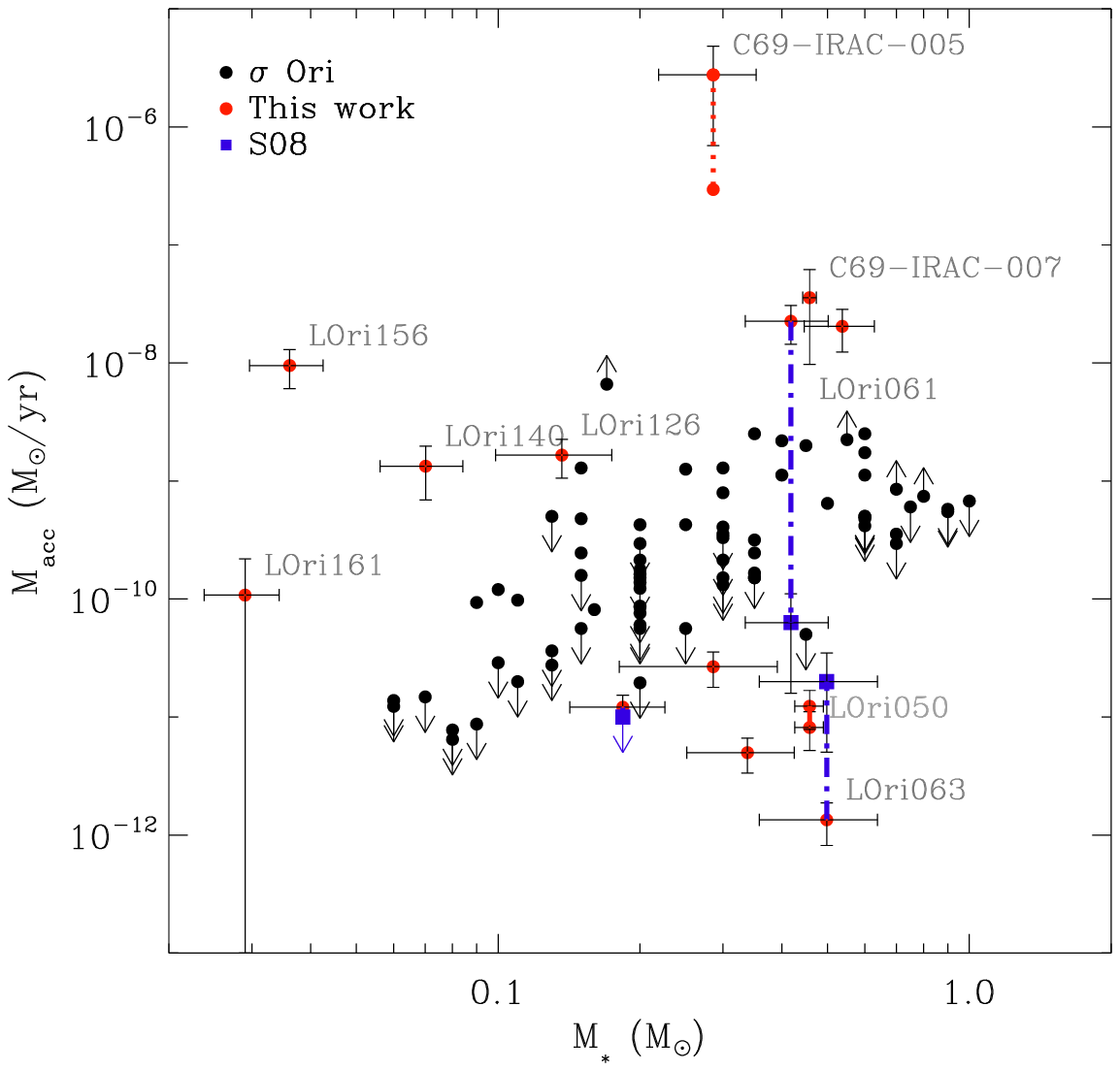}
\caption{Mass accretion rates versus mass of the central object for accreting objects. In red and blue we show members of C69 (red dots for measurements from this work and blue squares for those from \citealt{Sacco08}, the masses of LOri061 and LOri063 have been slightly shifted so that in the figure the comparison between the measurements of two studies is clearer) and in black members of the slightly younger cluster $\sigma$~Ori (measurements from \citealt{Rigliaco11}). We have included blue and red vertical bars to highlight special cases where more than one measurement is available for a given source (see Appendix~\ref{subsubsec:Haacc:ps} for details).
In all cases, the masses shown in this figure are the average between the one derived from the L$_{\rm bol}$ and the one derived from the T$_{\rm eff}$. The error bars display the differences among these determinations, see Paper I for details. Particular sources discussed in Appendix~\ref{subsubsec:Haacc:ps} are highlighted with grey labels.
}
\label{fig:M_Macc}
\end{figure}

The accretion rates calculated in this manner are also listed in Table~\ref{tab:paramTOTAL}. We note that for DM006, classified as accretor, we did not have a measurement of the FW$_{10\%}$ and therefore we could not estimate the accretion rate.

In Fig.~\ref{fig:M_Macc} we show an accretion rate versus mass diagram where we compare the values estimated in this work and those provided in \cite{Sacco08} also for C69; with those derived in \cite{Rigliaco11} for the slightly younger cluster $\sigma$ Orionis ($\sim$3 Myr according to their HR diagram). 

\cite{Rigliaco11} suggested in their work that there are two trends on this diagram with an inflection point at $\sim$0.45M$_{\odot}$. The mass range in C69 for which we have detected active accreting sources does not allow us to check this feature. Besides, the C69 sample is smaller and the dispersion of our measurements is much larger than that derived for $\sigma$ Orionis by \cite{Rigliaco11}. 

This larger dispersion can arise at a first stage from the different methodology used to estimate the mass accretion rate. While \cite{Sacco08} and this work use the H$\alpha$ emission, ``contaminated" by chromospheric activity as already discussed, \cite{Rigliaco11} use U-band photometry; a ``cleaner" methodology. On the other hand, C69 is likely older than $\sigma$ Orionis and the accretion disks of the former may be in a different evolutionary state than those of the latter.

Overall, our measurements are consistent with the idea that the accretion rate scales with the mass of the central object for low-mass stars. But, given the dispersion obtained, this is just a very rough trend. Individual sources from Figs.~\ref{fig:HaSpT} and~\ref{fig:M_Macc} are analyzed in Appendix~\ref{subsubsec:Haacc:ps}

Finally, to have a better understanding of the relation of the H$\alpha$ emission with the accretion process, we have tried to correlate that emission with disk properties derived mainly from the mid-infrared photometry. 

\begin{figure*}[htbp]
\begin{center}
\includegraphics[width=8.7cm]{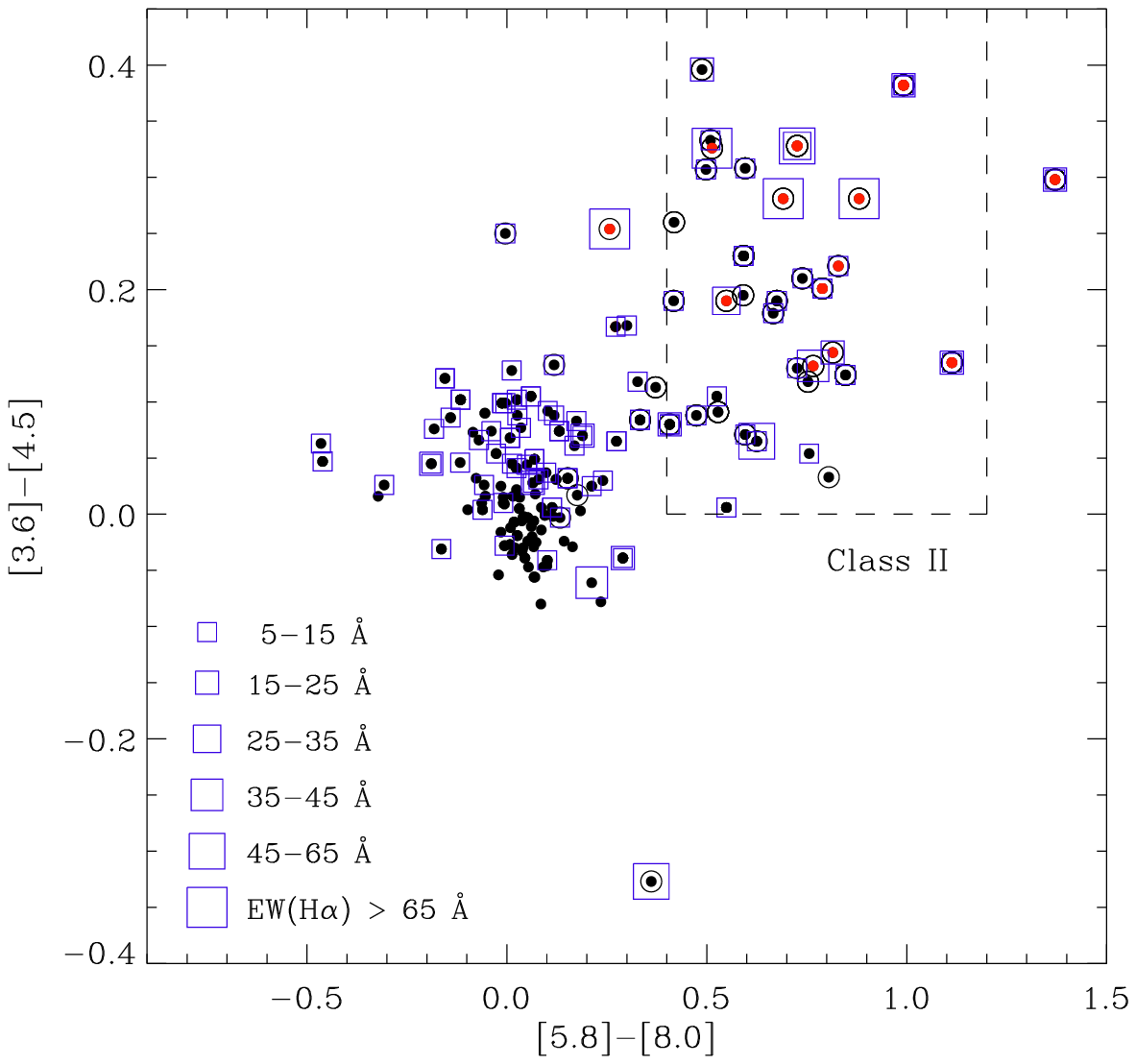}
\includegraphics[width=8.7cm]{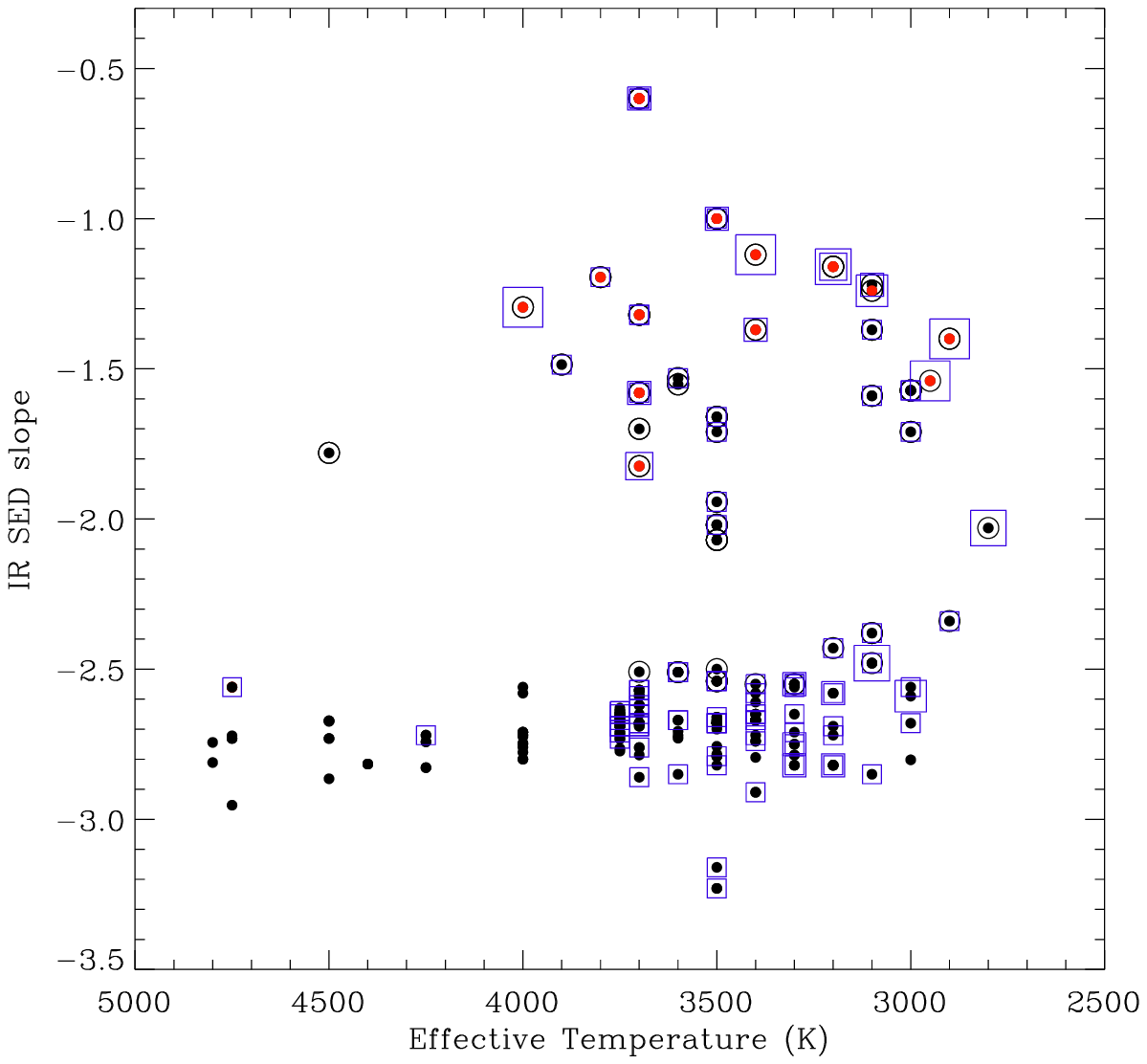}
\caption{Confirmed members with H$\alpha$ in emission compiled in Paper I. {\bf Left:} IRAC color-color diagram including information about the nature of the disk and the H$\alpha$ emission of the sources. Sources showing infrared excess have been surrounded by a black circumference. A blue square with a size proportional to the intensity of the H$\alpha$ emission of the sources is also shown in the figure. Whenever no square is present, the EW of the line is lower than 5\AA. The Class II area after \cite{Allen04} is highlighted with dashed lines.
{\bf Right:} Effective temperature vs. IRAC SED slope of the confirmed and candidate members of C69 including information about the nature of the disk and the H$\alpha$ emission of the sources as in the left panel.}
\label{fig:IRAC_ccd_acc}
\end{center}
\end{figure*}

The theoretical disk models used to interpret the IRAC [3.6]-[4.5] vs [5.8]-[8.0] color-color diagram by \cite{Allen04} suggest that the accretion rates increase from the bottom-left to the top-right of the Class II region, due to the increase of both the disk emission and the wall emission. In Fig.~\ref{fig:IRAC_ccd_acc} we show the mentioned IRAC color-color diagram for the members of C69 (spectroscopically confirmed members compilation from Paper I). We have included information regarding the presence of disks (large red circles), the intensity of the emission of H$\alpha$ (sized blue squares) and the classification as accretors (in red). 
The general trend agrees with the disk theory since objects with larger H$\alpha$ equivalent widths (up to accreting sources) have redder colors. A similar trend can be observed in the right panel of the same figure where we display the mid infrared SED slope as a function of effective temperature. Objects with optically thick disks seem to exhibit more intense H$\alpha$ emission. On the other hand, as it was already clear in Fig.~\ref{fig:HaSpT}, a large fraction of the sources ($\sim$65\%) harboring disks in C69 do not seem to be accreting from their disks. We will analyze this fact in more detail in section~\ref{subsec:holes}. 

\subsection{Disk with low H$\alpha$. Binaries clearing the inner disks?}
\label{subsec:holes}

\begin{figure}[htbp]
\begin{center}
\includegraphics[width=9cm]{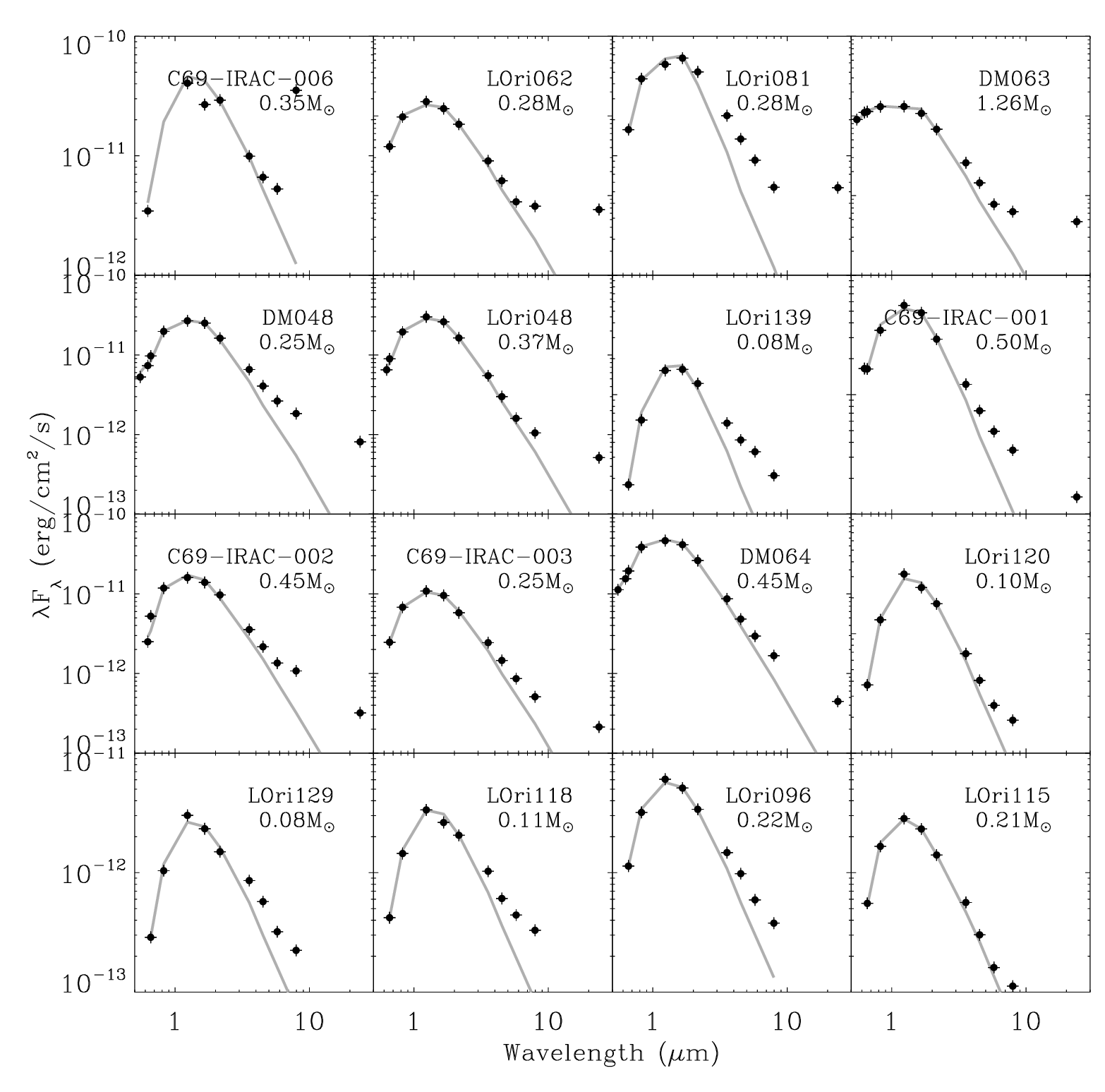}
\caption{SEDs of the sources classified as Class II based on their IRAC colors and showing EW(H$\alpha$) low enough not to be classified as accretors according to \cite{Barrado03}.}
\label{fig:SED_classII_not_ac}
\end{center}
\end{figure}

In the previous section we highlighted the presence of a large population of disk-bearing sources that do not show any signpost of active accretion.
In Fig.~\ref{fig:SED_classII_not_ac} we show examples of SEDs corresponding to this class of sources. We looked for some characteristic that would differentiate these objects from the others in our sample (other than the measured H$\alpha$ equivalent width). Their effective temperatures are mainly colder than $\sim$3750K, but, as explained in the next section, that is characteristic of all the disk population of C69. 

One possibility for these ``quiet" disks would be that they are dissipating their inner disks, but we see no difference in the near-infrared colors with respect to the actively accreting disks.

Besides, adopting the characterization of the $\alpha$ parameter from \cite{Lada06}, these objects harbor mainly optically thick disks (60\% of them), but there are also sources with optically thin disks and the so called transition/cold disks \citep{Merin10}. There are several mechanisms to explain the evolutionary status of these transition disks; one of the most attractive, in the context of planet formation, is the clearing of the inner disk by a giant planet in its earliest stages of formation (see the first observational candidate for this scenario in \citealt{Huelamo11}). Another possibility is tidal truncation in close binaries \citep{Ireland08}. This seems to be the mechanism at work in at least one of the cases of the low H$\alpha$ transition disks of C69; LOri043 that was classified as SB2 by \cite{Maxted08}.
On the other hand, LOri043, is the only documented spectroscopic binary in this sample of quiet disks. 

Finally, only two of the sources from this set have been detected in X-rays, suggesting that these are not particularly active objects either. In conclusion, we could not find any parameter (other than the H$\alpha$ emission) that unites these objects or differentiates them from the actively accreting population.

\section{Disk vs diskless populations}
\label{sec:distrib}

By the end of the previous section, we showed that we cannot trace differences between the actively accreting - quiet disk population of C69. In this section we will analyze the clear distinction between disk and diskless sources, in terms of their X-ray emission and their mass functions.

\subsection{H$\alpha$ and X-ray}
\label{subsec:HaXrays}

As stated in Section 2; \cite{Barrado11} presented the analysis of the XMM--Newton observations of two fields in C69. Several months later, \cite{Franciosini11} complemented the study by adding an extra field that covers the vicinity of the massive star $\lambda$ Ori, roughly at the center of the cluster.

These X-ray observations should trace well the weak-line T Tauri (and substellar analogs) population of C69; therefore we have combined the data from the two studies and correlate it with our census of spectroscopically confirmed members.

\begin{figure*}[htbp]
\begin{center}
\includegraphics[width=18cm]{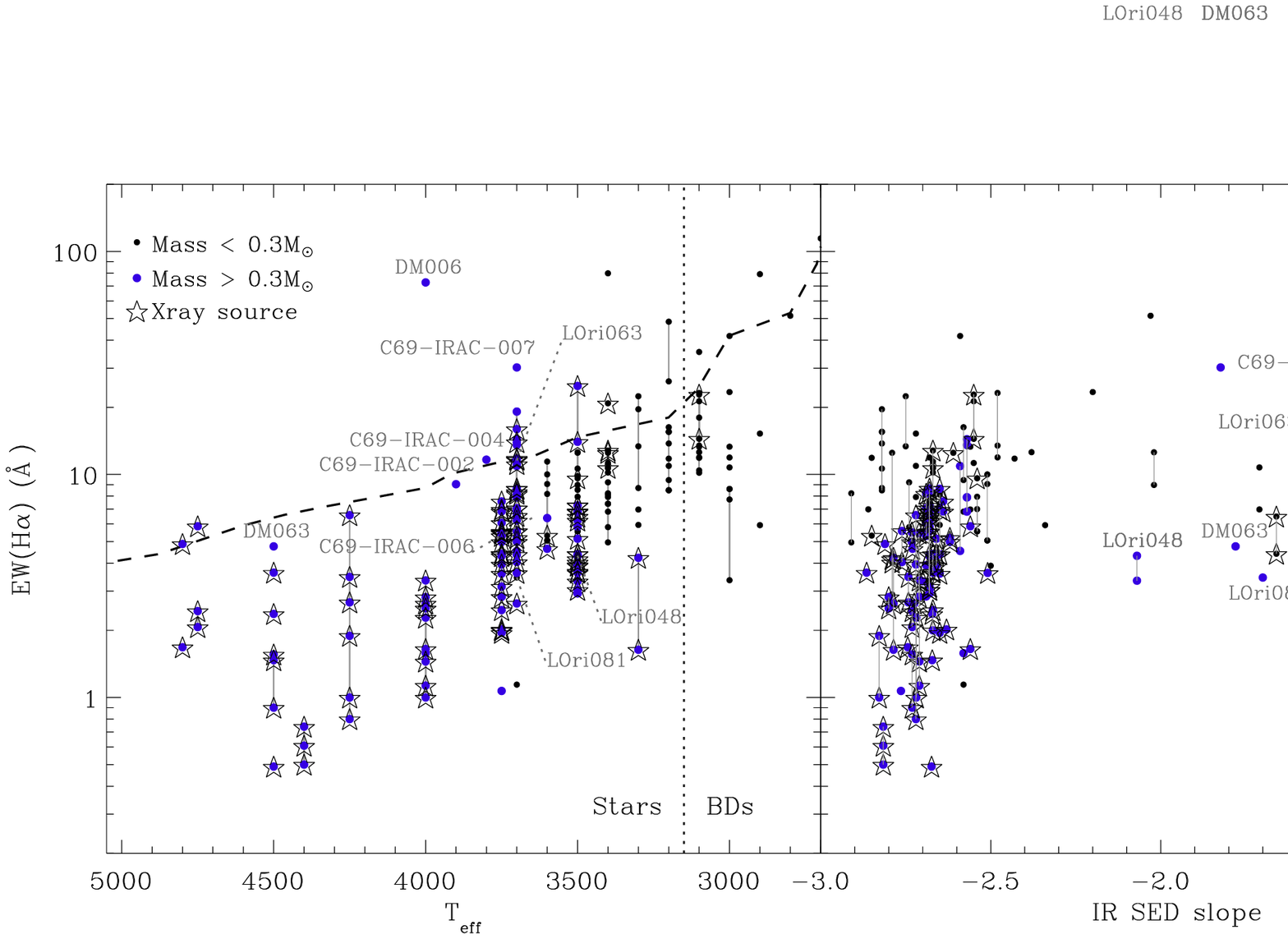}
\caption{Spectroscopically confirmed members from Paper I that lie in one of the two XMM-Newton pointings from \cite{Barrado11} or in the central one from \cite{Franciosini11}. We show in different colors objects with mass below (in black) and above (in blue) 0.3M$\odot$ (the completeness limit for \citealt{Barrado11}). For the EW(H$\alpha$) measurements we have gathered data from \cite{Sacco08, DM99} and this work. We have highlighted with large stars those sources detected in X-rays according to \cite{Barrado11} and/or \cite{Franciosini11}, and we have joined with a grey solid line different measurements of EW(H$\alpha$) corresponding to the same source. {\bf Left:} Effective temperature vs H$\alpha$ equivalent width (positive for emission) along with the saturation criterion. {\bf Right:} Mid-infrared slope vs H$\alpha$ equivalent width (same convention as in the previous panel)}
\label{fig:Halpha_Xrays}
\end{center}
\end{figure*}

In Fig.~\ref{fig:Halpha_Xrays} we illustrate this advantage of the X-ray observations to unveil the weak-line T Tauri population. We show every member of C69 lying in the field covered by XMM-Newton observations. We have highlighted in blue objects above the completeness limit of 0.3M$_{\odot}$. We see how most of the sources classified as weak-line T Tauri, according to the saturation criterion, are detected in X-rays. We also show that the objects that are not detected in X-rays are preferentially those harboring optically thick disks (both, active accretors and non-accreting sources). A total of 9 sources out of 12 with masses above 0.3M$_{\sun}$, harboring disks and within the XMM-Newton fields of view are not detected in X-rays and are labeled in Fig.~\ref{fig:Halpha_Xrays} 

\begin{figure}[htbp]
\begin{center}
\includegraphics[width=9cm]{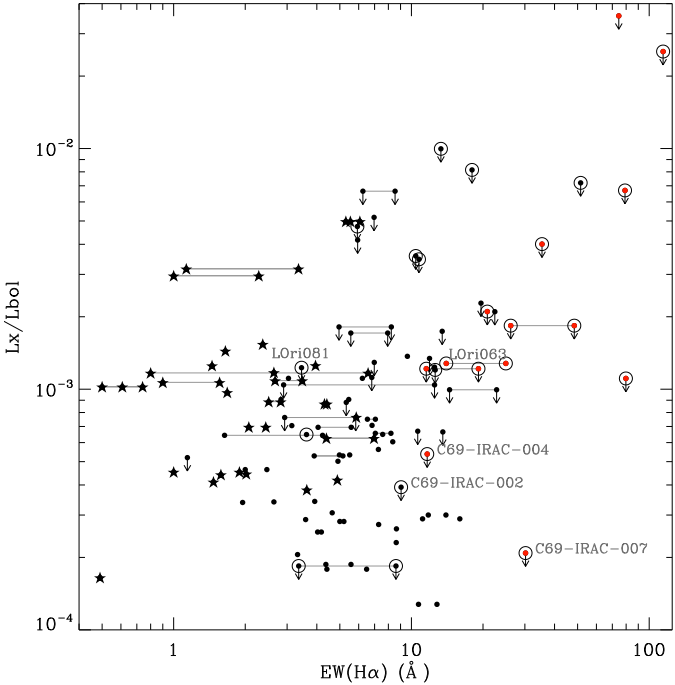}
\caption{EW(H$\alpha$) vs L$_{\rm X}$/L$_{\rm bol}$ diagram for the members from Paper I in the area covered in X-rays by either \cite{Barrado11} or \cite{Franciosini11}. As in previous figures we use large open circles to highlight objects showing mid-infrared excess and red dots for those classified as accretors. Five pointed stars indicate that the estimated mass of the object is larger than the critical mass 0.6M$_{\odot}$ explained in section~\ref{subsec:diskfrac}}
\label{fig:Halpha_Lx}
\end{center}
\end{figure}

This dichotomy is shown even clearer in Fig.~\ref{fig:Halpha_Lx}, where we show the X-ray luminosity vs bolometric luminosity ratio as a function of H$\alpha$ equivalent width. Here, we see how most objects with EW(H$\alpha$)  between $\sim$5--20~\AA~and not harboring disks are detected, while those showing infrared excess are not. We have highlighted five of the non-detected sources discussed in the previous paragraph for which \cite{Barrado11} provide upper-limits of the X-ray luminosity. The remaining four non-detections are located towards the center of the cluster, within the field of view of the observations by \cite{Franciosini11} and no upper-limit for the X-ray luminosity of the sources is provided in that work. 

\subsection{Disk fractions}
\label{subsec:diskfrac}

\begin{figure}[htbp]
\begin{center}
\includegraphics[width=9.cm]{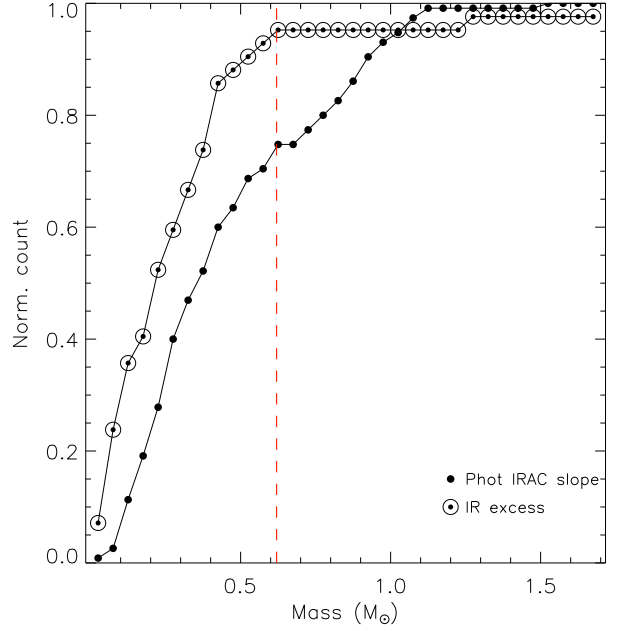}
\caption{Cumulative (normalized) mass functions for sources showing signs of harboring disks (infrared excess), plotted with large circles surrounding the points of the histogram, together with the cumulative mass function of sources detected by IRAC showing photospheric MIR slope. The vertical red dashed lines highlights the mass ($\sim$0.6M$_{\odot}$) at which the disk population is almost complete.}
\label{fig:mass_cum_func}
\end{center}
\end{figure}

Regarding the formation mechanism of brown dwarfs; if they are just a scaled down version of stars, one would think that the disk life-times above and below the hydrogen burning limit should be the same. On the other hand, there is
some evidence for longer life-times for infrared excesses in very low-mass stars and brown dwarfs than in higher-mass stars \citep{Lada06,Allers07}.

To test these possible differences, instead of estimating a typical star and brown dwarf disk fraction, we have divided our census of confirmed members into two sets:  sources showing some infrared excess, and objects detected also in the mid-infrared, but exhibiting purely photospheric colors. The first set contains 43 sources and the second one 115. We propose this approach to avoid choosing a fixed frontier in mass between stars and brown dwarfs so that our findings are easier to compare with other studies. Regarding the sources with excess, we have considered, in the same set, all kinds of MIR slopes and shapes:  optically thick, optically thin and transition disks.

We have computed the cumulative mass functions for both samples and the result is shown in Fig.~\ref{fig:mass_cum_func}: there is a clear difference at $\sim$0.6M$_{\odot}$ ($\sim$M2 spectral type). While the mass function of the diskless population rises up to $\sim$1.1M$_{\odot}$, the disk population is almost completely composed of sources with masses lower or equal than 0.6M$_{\odot}$. In other words: while we find diskless objects for every bin in mass, sources more massive than $\sim$0.6M$_{\odot}$ seem to have lost their disks already.
We tested the dependence of this change of behavior with the 5 Myr age assumed for C69 to estimate the masses. While an older age will significantly affect the mass determination of the lowest mass members of C69; that is not the case for sources with masses above 0.3M$_{\odot}$ according to the isochrones by \cite{Baraffe98, Chabrier00, Allard03}, and therefore our result is robust against changes in the age determination of C69.

\begin{figure}[htbp]
\begin{center}
\includegraphics[width=9.cm]{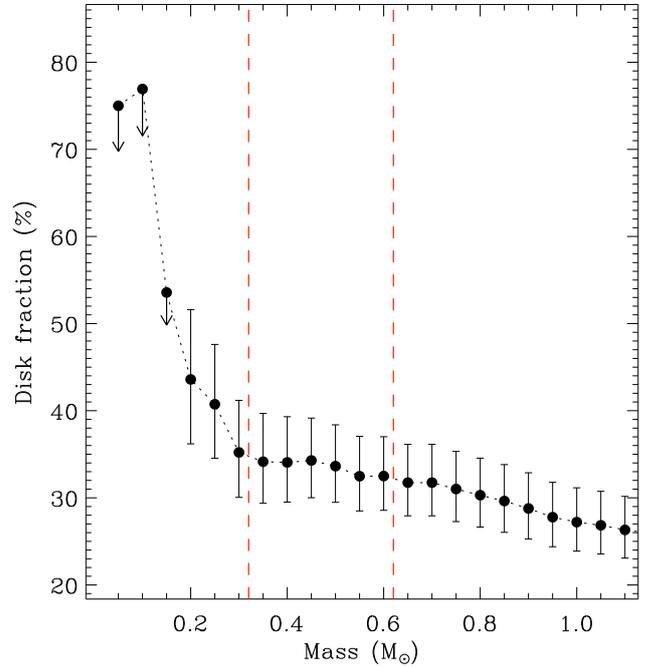}
\caption{Disk fraction as a function of the mass of the members of C69. For the ratio we use the same sets of objects as in the previous figure.}
\label{fig:disk_frac}
\end{center}
\end{figure}

On the other hand, if we use these two cumulative fractions but without normalization to estimate the disk fraction as a function of the mass, we can see that the situation is more complex. In Fig.~\ref{fig:disk_frac} and Table~\ref{tab:diskfraction} we show the ratio of the two cumulative functions; that is, for a given mass $M_i$, we provide $n_i/N_i$ where $n_i$ is the total number of sources with mass $\le M_i$ and infrared excess, and $N_i$ is the total number of sources with mass $\le M_i$ ($n_i$ plus the sources with purely photospheric infrared slopes and masses $\le M_i$).
For the error treatment, since our sample is very large, for masses larger or equal than 0.3M$_{\odot}$ we can derive standard Poisson uncertainty limits and for the lower masses we have used the approach described in \cite{Burgasser03}. 

Regarding the completeness of the diskless population, even though the X-ray observations are only complete down to 0.3M$_{\odot}$ we are confident that this does not induce any bias in our analysis. The reason is that every source that in the spectroscopic confirmation turned out to be later than M3.5 and was detected in X-rays, had previously already been selected as an optical photometric candidate. Therefore the optical photometric selection was as good as the X-rays one picking up the low to very low-mass members of C69.

The disk fraction function seems to peak (77\%) at the brown dwarf boundary, dropping abruptly with increasing mass up to $\sim$0.3M$_{\odot}$. It then stabilizes at $\sim$33\% before falling again for masses higher than $\sim$0.6M$_{\odot}$. The total stellar disk fraction is 26$^{+4}_{-3}$ \%

We must note that the extremely high disk fraction for substellar objects should be taken as an upper-limit. In this low mass regime, we have some sources that have not been detected in the two reddest channels of Spitzer/IRAC (5.8 and 8.0 $\mu$m). We did not consider those sources since we only used sources with photospheric MIR SED to estimate the disk fractions. If we were certain that those sources do not show excess (and the fact that they are not detected at those red wavelengths is a good indicator of that), the percentage would decrease down to $\sim$58\%

This fraction is still larger than that derived by \cite{Barrado07} for the same cluster ($\sim$40\%). This is not too surprising since in this study we consider the fraction with any kind of disk and not only Class II sources as in \cite{Barrado07}. On the other hand, the difference with the value derived by \cite{Scholz07} for Upper Sco ($\sim$37\%), a similar age cluster; although significant, could just be caused by small number statistics or again, use of different criterion to infer whether a source is harboring a disk or not. 

\begin{table}
\begin{center}
\caption{Disk fraction as a function of mass.}
\label{tab:diskfraction}
\begin{tabular}{lc|lc}
\hline\hline
$<$ M(M$_{\odot}$) & Disk frac (\%) & $<$ M(M$_{\odot}$) & Disk frac (\%) \\
\hline
0.1  & $\le$ 77              & 0.75 & 31$^{+4}_{-4}$  \\
0.15 & $\le$ 54             & 0.8  & 30$^{+4}_{-3}$  \\ 
0.2  & 44$^{+8}_{-7}$ & 0.85 & 30$^{+4}_{-4}$  \\ 
0.25 & 41$^{+7}_{-6}$& 0.9  & 29$^{+4}_{-4}$  \\ 
0.3  & 35$^{+6}_{-5}$ & 0.95 & 28$^{+4}_{-3}$  \\ 
0.35 & 34$^{+6}_{-5}$ & 1.0  & 27$^{+4}_{-3}$  \\ 
0.5  & 34$^{+5}_{-4}$ & 1.05 & 27$^{+4}_{-3}$  \\ 
0.55 & 32$^{+5}_{-4}$ & 1.1  & 26$^{+4}_{-3}$  \\ 
0.7  & 32$^{+4}_{-4}$ & 1.7  & 26$^{+4}_{-3}$  \\ 
\hline\hline
\end{tabular}
\end{center}
\end{table}

\subsection{Spatial distribution}

\begin{figure}[htbp]
\begin{center}
\includegraphics[width=9.cm]{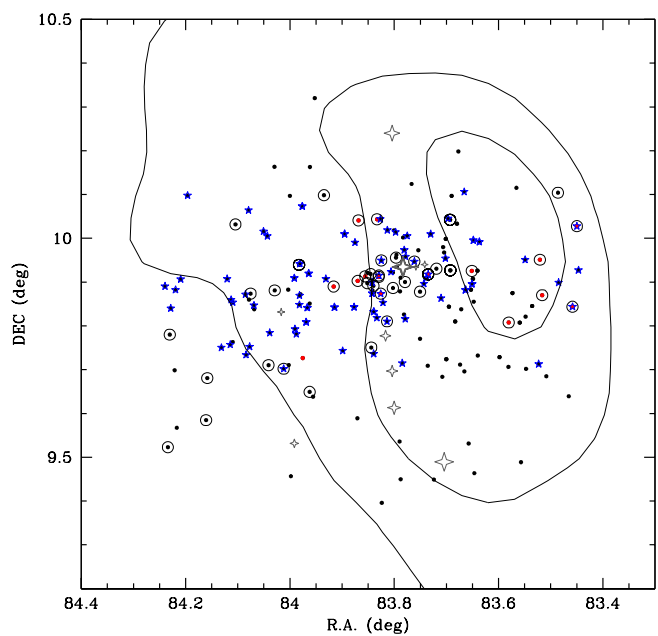}
\caption{Spatial distribution of the spectroscopically confirmed members of C69. Active accretors according to the saturation criterion are highlighted in red. Information regarding infrared excess (as a proxy for the presence of disk, from \citealt{Barrado07}) is provided by surrounding those sources with larger circumferences. The more massive population of C69 ($\lambda$ Ori itself and the B stars) is shown with grey four-pointed stars. Finally, the members with X-ray detections are shown with five-pointed blue stars.}
\label{fig:dist_Halpha}
\end{center}
\end{figure}

In Fig.~\ref{fig:dist_Halpha} we show the spatial distribution of C69 spectroscopically confirmed members (by \citealt{DM99,DM01,Sacco08, Maxted08} or this work) including information about X-ray emission (from \citealt{Barrado11} and/or \citealt{Franciosini11}), the presence of a disk \citep{Barrado07} and whether the disk is accreting or not.  As can be seen in the figure, the sources with a disk show a higher concentration towards the center of the cluster with respect to the diskless population (contrary to what one would expect according to the supernovae scenario and already suggested in \citealt{Barrado07}). In fact, if we assume that actively accreting systems are younger than those which do not show any sign of accretion, the youngest population of C69 seems to be clustered either around the central star $\lambda$ Ori or to the South-West.

We have computed the two-sided Kuiper statistic (invariant Kolmogorov-Smirnov), and the associated probability that any of the previously mentioned populations were drawn from the same distribution. 
The tests reveal that the cumulative distribution function of Class II candidates is very different from that of Class III, with a 99.9\% probability that both populations have been drawn from different distributions.

The conclusion from this test is that objects with masses lower than 0.6M$_{\odot}$ have been less efficient, in the life time of C69, in loosing their circumstellar material. 

\section{Summary and conclusions}
\label{sec:conclusions}

We have analyzed the spectroscopic properties of the very complete sample of members of C69 compiled in \cite{Bayo11}. Using different spectroscopic features we have tried to better understand the similarities and differences between the disk and diskless populations and the stellar and substellar ones. Our main results can be summarized as follows:


\begin{enumerate}

\item We have estimated the rotational velocities for eight members. We find a high dispersion in the v$\sin(i)$ values, being larger among the diskless population. We interpret this as a result of disk locking in some of the C69 members.

\item We have studied the variability of the H$\alpha$ emission line in 142 members classified as magnetically active (non-accreting sources). We have also identified candidates that might have experienced flares during the epoch of observation.

\item We have tried to disentangle the activity and accretion contributions to the H$\alpha$ emission. In this context, for those sources showing large H$\alpha$ equivalent widths (larger than the one expected to arise from chromospheric activity) we have derived accretion rates using mainly the full width at 10\% of the flux of the H$\alpha$ emission line (but also the Ca II IRT) obtaining a very large spread of values. Once individual cases are analyzed, these spread values are still compatible with those previously reported in the literature for objects of similar mass and age \citep{Rigliaco11}. When considering all the confirmed members with H$\alpha$ measurements, we estimate a 9$^{+3}_{-2}$ \% fraction of accretors in C69. If we only consider objects with disks, 38$^{+8}_{-7}$ \% show active accretion.

\item We have studied the relation H$\alpha$ -- disk properties. While the general trend expected from disk models applies, we also identify a pretty large population of ``quiet" disks. Objects showing clear mid-infrared excess (with a variety of SED shapes) but H$\alpha$ levels compatible with arising from pure chromospheric activity and not from the interaction with the disk via accretion.

\item We find a stellar disk fraction of 26$^{+4}_{-3}$ \% and we can put some limits to the substellar disk fraction (for the faintest members, due to sensitivity limitations, the lack of detection at 5.8 and/or 8 micron does not imply the presence or absence of a substellar disk) of 58\%. These fractions do not compare too badly (taking into account the difference in procedure followed to estimate the fraction) with that previously provided for C69 itself by \cite{Barrado04} ($\sim$40\%) and is significantly higher than that derived by \cite{Scholz07} for the cluster of similar age Upper Sco ($\sim$37\%). 

\item Regarding the accretion fraction in the substellar domain, we do not see dramatic changes from the global fraction of C69. This accretion fraction for brown dwarfs varies from 30\% to $\sim$43\% according to our uncertainties in mass determination, This range is also compatible with the one provided by \cite{Scholz07} for the substellar population of Upper Sco.

\item We have confirmed that X-ray observations \citep{Barrado11, Franciosini11} are extremely efficient recovering the Class III population of C69 in the intermediate to low-mass range.

\item We have compared the mass function of the disk and diskless populations of C69 finding that 0.6M$_{\odot}$ seems to be the critical mass below which a significant fraction of the members preserve their disks. This result implies different disk lifetimes for different stellar masses. In particular, for masses lower than 0.6M$_{\odot}$ we have shown that the disk fraction rises very steeply with the caveat that in the brown dwarf domain the fraction provided should be taken as an upper-limit.

\item We have studied the spatial distribution of the disk-harboring population of C69 and we find that, opposite to what \cite{DM01} derived from their more massive members sample; the density of disk-sources is larger closer to the center of the cluster, which is inconsistent with the SN scenario invoked to explain the origin of the Lambda Orionis star forming region. In addition, the winds (current or in the recent past) from the massive star $\lambda$ Ori, seem not to have affected the distribution of disk and diskless cluster members.

\end{enumerate}

\begin{acknowledgements}
A. Bayo would like to thank B. Montesinos for the interesting discussion about the rotational velocities determination and H. Bouy and M. Lopez del Fresno for very useful advices in statistics. This publication makes use of VOSA, developed under the Spanish Virtual Observatory project supported from the Spanish MICINN through grant AYA2008-02156. This work was co-funded under the Marie Curie Actions of the European Commission (FP7-COFUND) and Spanish grants AYA2010-21161-C02-02, CDS2006-00070 and PRICIT-S2009/ESP-1496.

\end{acknowledgements}


\bibliographystyle{aa}
\bibliography{./biblio}

\onllongtab{6}{
\begin{landscape}
\tiny
\begin{longtable}{lcllclclll}
\caption{Parameters and measurements obtained in this work for Collinder 69 confirmed members and comparison with previous studies.\label{tab:paramTOTAL}}\\
\hline\hline
Object & Bin$^1$ & Class$^2$ &Disk Type &$v$sin$i$$^3$& Acc.$^4$ & log(Macc) & EW(H$\alpha$)$^5$ & FW$_{10\%}$(H$\alpha$)$^5$ & EW(Ca II) \\
       &         &           &          &(km/s)       &          & (log(M$_{\odot}$/yr))&  (\AA) & (km/s)                       &  (\AA) \\      
       &         &           &          &             &          &           &     6563 \AA      &                            & 8498, 8542, 8662 \AA \\      
\hline
\endfirsthead

\multicolumn{5}{c}{{\tablename} \thetable{} -- \scriptsize Continued} \\
\hline\hline
Object & Bin$^1$ & Class$^2$ &Disk Type &$v$sin$i$$^3$& Acc.$^4$ & log(Macc) & EW(H$\alpha$)$^5$ & FW$_{10\%}$(H$\alpha$)$^5$ & EW(Ca II) \\
       &         &           &          &(km/s)       &          & (log(M$_{\odot}$/yr))&  (\AA) & (km/s)                       &  (\AA) \\      
       &         &           &          &             &          &           & 6563 \AA          &                            & 8498, 8542, 8662 \AA \\      
\hline
\endhead

\hline \multicolumn{5}{r}{{Continued on next page\ldots}} \\ \hline
\endfoot

\hline\hline
\endlastfoot
 LOri001      &               & III & Diskless   &                       & N &                  & -2.82 $\pm$0.16$^{1)}$, -2.51$^{2)}$                           &                                                 &                       \\
 LOri003      & SB (S08)      & III & Diskless   &                       & N &                  &                         -3.35$^{2)}$  , -1.13 $\pm$0.15$^{3)}$ &                                                 &                       \\
 LOri008      &               & III & Diskless   &                       & N &                  &                         -1.65$^{2)}$                           &                                                 &                       \\
 LOri013      &               & III & Diskless   &                       & N &                  & -4.31 $\pm$0.21$^{1)}$, -4.41$^{2)}$                           &                                                 &                       \\
 LOri014      &               & III & Diskless   &                       & N &                  &                         -1.45$^{2)}$                           &                                                 &                       \\
 LOri016      & SB (S08)      & III & Diskless   &                       & N &                  &                                         -1.58 $\pm$0.13$^{3)}$ &                                                 &                       \\
 LOri017      &               & III & Diskless   & 70$^{*)}$             & N &                  & -6.57 $\pm$2.31$^{1)}$, -0.8 $^{2)}$                           &                                                 &                       \\
 LOri018      &               & III & Diskless   &                       & N &                  &                         -2.02$^{2)}$                           &                                                 &                       \\
 LOri022      &               & III & Diskless   & $<$17.0               & N &                  &                         -4.39$^{2)}$  , -6.96 $\pm$0.69$^{3)}$ &                                                 &                       \\
 LOri023      &               & III & Diskless   &                       & N &                  &                         -1.95$^{2)}$                           &                                                 &                       \\
 LOri024      &               & III & Diskless   & $<$17.0               & N &                  & -2.47 $\pm$0.11$^{1)}$,                 -2.0  $\pm$0.12$^{3)}$ &                                                 &                       \\
 LOri025      &               & III & Diskless   &                       & N &                  &                         -3.95$^{2)}$                           &                                                 &                       \\
 LOri026      &               & III & Diskless   & 26.0$^{+6.0}_{-1.0}$  & N &                  & -5.54 $\pm$0.30$^{1)}$, -6.07$^{2)}$  , -5.3  $\pm$0.3 $^{3)}$ &                                                 &                       \\
 LOri030      & SB2 (S08)     & III & Diskless   &                       &   &                  &                                                                &                                                 &                       \\
 LOri031      &               & III & Diskless   & 40$^{*)}$             & N &                  & -2.84 $\pm$0.14$^{1)}$                                         &                                                 &                       \\
 LOri032      &               & III & Diskless   &                       & N &                  &                         -6.83$^{2)}$                           &                                                 &                       \\
 LOri033      &               & III & Diskless   &                       & N &                  &                         -3.14$^{2)}$                           &                                                 &                       \\
 LOri034      &               & II  & Thick      & $<$17.0               & Y &                  &                                         -8.47 $\pm$0.47$^{3)}$ &                               215$\pm$25$^{3)}$ &                       \\
 LOri035      &               & III & Diskless   & 30.2$^{+10.0}_{-6.2}$ & N &                  &                                         -3.32 $\pm$0.35$^{3)}$ &                                                 &                       \\
 LOri037      &               & III & Diskless   & $<$17.0               & N &                  & -4.40 $\pm$0.23$^{1)}$, -3.63$^{2)}$  , -3.67 $\pm$0.34$^{3)}$ &                                                 &                       \\
 LOri038      &               & II  & Thick      &                       & Y & -10.90$\pm$0.03  & -14.01$\pm$0.51$^{1)}$,-24.95$^{2)}$                           & 164$\pm$26$^{1)}$                               &                       \\
 LOri039      &               & III & Diskless   &                       & N &                  &                         -3.59$^{2)}$                           &                                                 &                       \\
 LOri040      &               & III & Diskless   & $<$21.3               & N &                  &                         -3.9 $^{2)}$  , -5.15 $\pm$0.62$^{3)}$ &                                                 &                       \\
 LOri041      &               & III & Diskless   & 61.9$^{+11.8}_{-5.7}$ & N &                  &                         -8.2 $^{2)}$  , -7.05 $\pm$0.86$^{3)}$ &                                                 &                       \\
 LOri042      &               & III & Diskless   & 30$^{*)}$             & N &                  & -3.02 $\pm$0.16$^{1)}$                                         &                                                 &                       \\
 LOri043      & SB2 (M08)     & III & Transition &                       & N &                  & -3.92 $\pm$0.12$^{1)}$                                         &                                                 &                       \\
 LOri045      &               & III & Diskless   & $<$17.0               & N &                  & -6.85 $\pm$0.11$^{1)}$,                 -3.04 $\pm$0.33$^{3)}$ &                                                 &                       \\
 ``           &               & III & Diskless   & $<$17.0               & N &                  & -6.22 $\pm$0.34$^{1)}$                                         &                                                 &                       \\
 LOri047      &               & III & Diskless   &                       & N &                  &                         -8.65$^{2)}$                           &                                                 &                       \\
 LOri048      &               & II  & Thin       & $<$17.0               & N &                  & -4.32 $\pm$0.19$^{1)}$,                 -3.34 $\pm$0.44$^{3)}$ &                                                 &                       \\
 LOri050      & SB1(M08, S08) & II  & Thick      & 60$^{*)}$             & Y & -11.09$\pm$0.05  & -11.65$\pm$1.29$^{1)}$,                 -11.14$\pm$0.96$^{3)}$ & 186$\pm$37$^{1)}$ ,                             &                       \\
 ``           & SB1(M08 S08)  & II  & Thick      &                       & Y & -10.91$\pm$0.05  & -15.98$\pm$1.58$^{1)}$,                                        & 204$\pm$ 5$^{1)}$ ,                             &                       \\
 LOri051      & SB1 (M08)     & III & Diskless   &                       &   &                  &                                                                &                                                 &                       \\
 LOri052      &               & III & Diskless   &                       & N &                  & -6.80 $\pm$0.44$^{1)}$                                         &                                                 &                       \\
 ``           &               & III & Diskless   &                       & N &                  & -7.26 $\pm$0.42$^{1)}$                                         &                                                 &                       \\
 LOri053      &               & III & Diskless   & $<$18.3               & N &                  & -4.30 $\pm$0.15$^{1)}$,                 -2.93 $\pm$0.2 $^{3)}$ &                                                 &                       \\
 ``           &               & III & Diskless   & $<$18.3               & N &                  & -5.83 $\pm$0.22$^{1)}$,                                        &                                                 &                       \\
 LOri054      &               & III & Diskless   &                       & N &                  & -8.35 $\pm$0.48$^{1)}$                                         &                                                 &                       \\
 LOri055      &               & III & Diskless   & $<$17.0, $<$20$^{*)}$ & N &                  & -7.05 $\pm$0.13$^{1)}$,                 -6.53 $\pm$0.49$^{3)}$ &                                                 &                       \\
 LOri056      &               & III & Diskless   & $<$19.3               & N &                  & -6.22 $\pm$0.22$^{1)}$,                 -5.57 $\pm$0.81$^{3)}$ &                                                 &                       \\
 ``           &               & III & Diskless   & $<$19.3               & N &                  & -4.37 $\pm$0.45$^{1)}$,                                        &                                                 &                       \\
 LOri057      &               & III & Diskless   & $<$17.0, $<$20$^{*)}$ & N &                  & -5.00 $\pm$0.31$^{1)}$,                 -5.2  $\pm$0.52$^{3)}$ &                                                 &                       \\
 LOri058      &               & III & Diskless   &                       & N &                  & -7.28 $\pm$0.40$^{1)}$                                         &                                                 &                       \\
 LOri059      &               & III & Diskless   &                       & N &                  & -9.83 $\pm$0.30$^{1)}$                                         &                                                 &                       \\
 ``           &               & III & Diskless   & 20--40$^{*)}$         & N &                  & -6.87 $\pm$0.29$^{1)}$                                         &                                                 &                       \\
 LOri060      &               & III & Diskless   & $<$18.5               & N &                  & -4.03 $\pm$0.21$^{1)}$,                 -4.18 $\pm$0.48$^{3)}$ &                                                 &                       \\
 LOri061      &               & II  & Thick      & 17.9$^{+1.4}_{-1.8}$  & Y & -7.65$\pm$0.05   & -11.78$\pm$0.17$^{1)}$,                 -13.97$\pm$1.38$^{3)}$ & 540$\pm$ 5$^{1)}$ ,  274$\pm$15$^{3)}$          &                       \\
 LOri062      &               & II  & Thick      & $<$17.0               & N &                  & -6.48 $\pm$0.73$^{1)}$,                 -4.41 $\pm$0.3 $^{3)}$ &                                                 &                       \\
 LOri063      &               & II  & Thick      & $<$18.2   & Y &-11.87$\pm$0.07$^{1)}$, -10.7$\pm$0.3$^{3)}$& -11.55$\pm$0.69$^{1)}$,                 -19.14$\pm$2.53$^{3)}$ &                                                 &                       \\
 LOri064      &               & III & Diskless   &                       & N &                  & -9.62 $\pm$0.56$^{1)}$                                         &                                                 &                       \\
 LOri065      &               & III & Transition &                       &   &                  &                                                                &                                                 &                       \\
 LOri066      &               & III & Diskless   &                       &   &                  &                                                                &                                                 &                       \\
 LOri067      &               & III & Diskless   &                       &   &                  &                                                                &                                                 &                       \\
 LOri068      &               & III & Diskless   & $<$17.0               & N &                  & -14.39$\pm$0.74$^{1)}$,                 -7.9  $\pm$1.26$^{3)}$ &                                                 &                       \\
 ``           &               & III & Diskless   & $<$17.0               & N &                  & -6.82 $\pm$0.52$^{1)}$,                                        &                                                 &                       \\
 LOri069      & SB2(M08, S08) & III & Diskless   &                       & N &                  & -8.12 $\pm$0.47$^{1)}$,                 -7.39 $\pm$1.46$^{3)}$ &                                                 &                       \\
 LOri070      &               & III & Diskless   &                       & N &                  & -13.56$\pm$3.26$^{1)}$                                         &                                                 &                       \\
 LOri071      &               & III & Diskless   &                       & N &                  & -6.81 $\pm$0.23$^{1)}$                                         &                                                 &                       \\
 LOri072      &               & III & Diskless   &                       & N &                  & -11.26$\pm$0.51$^{1)}$                                         &                                                 &                       \\
 LOri073      &               & III & Diskless   &                       & N &                  & -4.54 $\pm$0.19$^{1)}$                                         &                                                 &                       \\
 ``           &               & III & Diskless   &                       & N &                  & -10.91$\pm$1.04$^{1)}$                                         &                                                 &                       \\
 LOri074      &               & III & Diskless   &                       &   &                  &                                                                &                                                 &                       \\
 LOri075      & SB1 (M08)     & III & Diskless   & 61.3$^{+11.5}_{-4.9}$ & N &                  & -10.20$\pm$0.89$^{1)}$,                 -10.71$\pm$0.98$^{3)}$ &                                                 &                       \\
 ``           & SB1 (M08)     & III & Diskless   & 65$^{*)}$             & N &                  & -10.73$\pm$0.73$^{1)}$,                                        &                                                 &                       \\
 ``           & SB1 (M08)     & III & Diskless   &                       & N &                  & -12.81$\pm$0.86$^{1)}$,                                        &                                                 &                       \\
 LOri076      &               & III & Diskless   & $<$17.0               & N &                  &                                         -4.92 $\pm$0.6 $^{3)}$ &                                                 &                       \\
 LOri077      &               & III & Diskless   &                       & N &                  & -7.92 $\pm$0.23$^{1)}$                                         &                                                 &                       \\
 LOri078      &               & III & Diskless   &                       & N &                  & -1.14 $\pm$0.19$^{1)}$                                         &                                                 &                       \\
 LOri079      &               & III & Diskless   & $<$17.0               & N &                  & -10.01$\pm$0.24$^{1)}$,                 -5.06 $\pm$0.53$^{3)}$ &                                                 &                       \\
 ``           &               & III & Diskless   & $<$17.0               & N &                  & -9.06 $\pm$3.35$^{1)}$,                                        &                                                 &                       \\
 LOri080      &               & III & Diskless   & 60.2$^{+21.3}_{-5.9}$ & N?& -10.37$\pm$0.05  & -21.3$\pm$1.00$^{1)}$,                  -14.47$\pm$1.25$^{3)}$ & 259$\pm$24$^{1)}$ ,  287$\pm$41$^{3)}$          &                       \\
 ``           &               & III & Diskless   & 60.2$^{+21.3}_{-5.9}$ &   &  -9.97$\pm$0.05  & -22.82$\pm$0.62$^{1)}$                                         & 301$\pm$25$^{1)}$                               &                       \\
 LOri081      &               & II  & Thick      &                       & N &                  & -3.45 $\pm$0.07$^{1)}$                                         &                                                 &                       \\
 LOri082      &               & III & Diskless   &                       & N &                  & -10.63$\pm$2.09$^{1)}$                                         &                                                 &                       \\
 LOri083      &               & III & Diskless   & 19.1$^{+7.9}_{-2.6}$  & N &                  &                                         -5.32 $\pm$0.68$^{3)}$ &                                                 &                       \\
 LOri084      &               & III & Diskless   &                       & N &                  & -10.42$\pm$3.18$^{1)}$                                         &                                                 &                       \\
 LOri085      &               & II  & Thick      &                       &   &                  &                                                                &                                                 &                       \\
 LOri086      &               & III & Diskless   &                       & N &                  & -6.98 $\pm$0.21$^{1)}$                                         &                                                 &                       \\
 LOri087      &               & III & Diskless   & 18.4$^{+6.0}_{-1.7}$  & N &                  & -7.00 $\pm$0.39$^{1)}$,                 -5.56 $\pm$0.7 $^{3)}$ &                                                 &                       \\
 ``           &               & III & Diskless   & 18.4$^{+6.0}_{-1.7}$  & N &                  & -7.95 $\pm$0.23$^{1)}$,                                        &                                                 &                       \\
 LOri088      &               & III & Diskless   & $<$17.0               & N &                  & -16.29$\pm$0.48$^{1)}$,                 -9.44 $\pm$1.12$^{3)}$ &                                                 &                       \\
 LOri089      &               & III & Diskless   &                       & N &                  & -3.90 $\pm$0.23$^{1)}$                                         &                                                 &                       \\
 LOri090      &               & III & Diskless   &                       &   &                  &                                                                &                                                 &                       \\
 LOri091      &               & III & Diskless   &                       & N &                  & -11.93$\pm$1.12$^{1)}$                                         &                                                 &                       \\
 ``           &               & III & Diskless   &                       & N &                  & -23.2$\pm$12.3$^{1)}$                                          &                                                 &                       \\
 ``           &               & III & Diskless   &                       & N &                  & -13.47$\pm$1.11$^{1)}$                                         &                                                 &                       \\
 LOri092      &               & III & Diskless   & 19.8$^{+2.8}_{-2.6}$  & N &                  & -12.51$\pm$1.37$^{1)}$,                 -2.9  $\pm$0.34$^{3)}$ &                                                 &                       \\
 LOri093      &               & III & Diskless   & $<$18.3               & N &                  & -8.54 $\pm$0.22$^{1)}$,                 -6.25 $\pm$0.94$^{3)}$ &                                                 &                       \\
 LOri094      &               & III & Diskless   & 54.8$^{+5.5}_{-8.2}$  & N &                  & -8.49 $\pm$0.10$^{1)}$,                 -15.53$\pm$2.23$^{3)}$ &                                                 &                       \\
 ``           &               & III & Diskless   & 54.8$^{+5.5}_{-8.2}$  & N &                  & -13.78$\pm$0.73$^{1)}$,                                        &                                                 &                       \\
 LOri095      &               & III & Diskless   & $<$19.7               & N &                  & -8.23 $\pm$1.18$^{1)}$,                 -4.96 $\pm$0.62$^{3)}$ &                                                 &                       \\
 LOri096      &               & II  & Thick      & $<$19.1               & N &                  &                                         -6.97 $\pm$1.1 $^{3)}$ &                                                 &                       \\
 LOri098      &               & III & Diskless   &                       & N &                  & -12.51$\pm$0.90$^{1)}$                                         &                                                 &                       \\
 LOri099      &               & III & Diskless   &                       & N &                  & -5.81 $\pm$0.64$^{1)}$                                         &                                                 &                       \\
 ``           &               & III & Diskless   &                       & N &                  & -9.22 $\pm$0.67$^{1)}$                                         &                                                 &                       \\
 LOri100      &               & III & Diskless   & $<$17.0               & N &                  & -11.43$\pm$1.84$^{1)}$,                 -8.18 $\pm$1.4 $^{3)}$ &                                                 &                       \\
 LOri102      &               & III & Diskless   & $<$17.0               & N &                  &                                         -5.94 $\pm$0.96$^{3)}$ &                                                 &                       \\
 LOri103      &               & III & Thin       &                       &   &                  &                                                                &                                                 &                       \\
 LOri104      &               & II  & Thick      &                       &   &                  &                                                                &                                                 &                       \\
 LOri105      &               & III & Diskless   & $<$20.0               & N &                  & -13.38$\pm$2.60$^{1)}$                                         &                                                 &                       \\
 ``           &               & III & Diskless   & $<$20.0               & N &                  & -22.43$\pm$0.45$^{1)}$                                         &                                                 &                       \\
 LOri106      &               & II  & Thick      & $<$17.0               & Y & -10.92$\pm$0.01  & -48.42$\pm$1.85$^{1)}$,                 -26.16$\pm$2.4 $^{3)}$ & 203$\pm$ 1$^{1)}$ ,  107$\pm$4 $^{3)}$          &                       \\
 LOri107      &               & III & Diskless   &                       & N &                  & -11.92$\pm$1.02$^{1)}$                                         &                                                 &                       \\
 LOri109      &               & III & Diskless   &                       & N &                  & -8.70 $\pm$0.45$^{1)}$                                         &                                                 &                       \\
 ``           &               & III & Diskless   &                       & N &                  & -19.62$\pm$0.52$^{1)}$                                         &                                                 &                       \\
 LOri112      &               & III & Diskless   &                       & N &                  & -10.93$\pm$1.68$^{1)}$                                         &                                                 &                       \\
 LOri113      &               & II  & Thick      &                       & Y & -10.57$\pm$0.03  & -20.85$\pm$0.91$^{1)}$                                         & 239$\pm$ 3$^{1)}$                               &                       \\
 LOri114      &               & II  & Thin       &                       & N &                  & -12.59$\pm$1.73$^{1)}$                                         &                                                 &                       \\
 LOri115      &               & II  & Thin       &                       & N &                  & -8.97 $\pm$0.88$^{1)}$                                         &                                                 &                       \\
 ``           &               & II  & Thin       &                       & N &                  & -12.58$\pm$0.59$^{1)}$                                         &                                                 &                       \\
 LOri116      &               & III & Diskless   &                       & N &                  & -11.78$\pm$1.54$^{1)}$                                         &                                                 &                       \\
 LOri117      &               & --  & Thin       &                       & N &                  & -23.42$\pm$1.44$^{1)}$                                         &                                                 &                       \\
 LOri118      &               & II  & Thick      &                       & N &                  & -10.43$\pm$0.21$^{1)}$                                         &                                                 &                       \\
 LOri119      &               & III & Diskless   &                       & N &                  & -11.88$\pm$0.41$^{1)}$                                         &                                                 &                       \\
 LOri120      &               & II  & Thick      &                       & N &                  & -10.18$\pm$1.97$^{1)}$                                         &                                                 &                       \\
 ``           &               & II  & Thick      &                       & N &                  & -13.30$\pm$1.0 $^{1)}$                                         &                                                 &                       \\
 LOri122      &               & III & Diskless   &                       &   &                  &                                                                &                                                 &                       \\
 LOri124      &               & III & Diskless   &                       & N &                  & -6.97 $\pm$1.78$^{1)}$                                         &                                                 &                       \\
 LOri125      &               & III & Diskless   &                       &   &                  &                                                                &                                                 &                       \\
 LOri126      &               & II  & Thick      &                       & Y & -8.78$\pm$0.10   & -35.42$\pm$3.50$^{1)}$                                         & 424$\pm$10$^{1)}$                               &                       \\
 LOri129      &               & II  & Thick      &                       & N &                  & -10.76$\pm$0.94$^{1)}$                                         &                                                 &                       \\
 LOri130      &               & III & Diskless   &                       & N &                  & -8.51 $\pm$0.63$^{1)}$                                         &                                                 &                       \\
 LOri131      &               & II  & Thin       &                       &   &                  &                                                                &                                                 &                       \\
 LOri134      &               & III & Thin?      &                       & N &                  & -5.93 $\pm$0.31$^{1)}$                                         &                                                 &                       \\
 LOri135      &               & III & Diskless   &                       & N &                  & -13.32$\pm$3.53$^{1)}$                                         &                                                 &                       \\
 LOri139      &               & II  & Thick      &                       & N &                  & -17.99$\pm$1.02$^{1)}$                                         &                                                 &                       \\
 LOri140      &               & II  & Thick      &                       & Y & -8.88$\pm$0.14   & -79.14$\pm$9.44$^{1)}$                                         & 414$\pm$14$^{1)}$                               &                       \\
 LOri143      &               & III & Diskless   &                       & N &                  & -41.78$\pm$4.66$^{1)}$                                         &                                                 &                       \\
 LOri146      &               & III & Thin       &                       &   &                  &                                                                &                                                 &                       \\
 LOri150      &               & --  & Diskless   &                       & N &                  & -15.25$\pm$2.05$^{1)}$                                         &                                                 &                       \\
 LOri155      &               & III & Thin       &                       & N &                  &-51.49$\pm$18.70$^{1)}$                                         &                                                 &                       \\
 LOri156      &               & III & Thick      &                       & Y & -8.02$\pm$0.10   &-114.46$\pm$1.81$^{1)}$                                         & 502$\pm$11$^{1)}$                               &                       \\
 LOri161      &               & --  &            &                       & Y?& -9.97$\pm$0.42   &-74.51$\pm$33.07$^{1)}$                                         & 301$\pm$43$^{1)}$                               &                       \\
 DM003        &               & III & Diskless   &                       & N &                  &                         -3.18$^{2)}$                           &                                                 &                       \\
 DM005        &               & III & Diskless   &                       & N &                  &                         -0.06$^{2)}$                           &                                                 &                       \\
 DM006        &               & II  & Thick      &                       & Y &                  &                        -72.64$^{2)}$                           &                                                 &                       \\
 DM007        &               & III & Diskless   &                       & N &                  &                          0.12$^{2)}$                           &                                                 &                       \\
 DM008        &               & III & Diskless   &                       & N &                  &                          0.09$^{2)}$                           &                                                 &                       \\
 DM009        &               & III & Diskless   & 18.9$^{+2.7}_{-1.9}$  & N &                  & -0.61 $\pm$0.04$^{1)}$, -0.74$^{2)}$  , -0.5$\pm$0.03  $^{3)}$ &                                                 &                       \\
 DM010        &               & III & Diskless   &                       & N &                  &                         -5.97$^{2)}$                           &                                                 &                       \\
 DM013        &               & III & Diskless   &                       & N &                  &                         -1.29$^{2)}$                           &                                                 &                       \\
 DM014        &               & III & Diskless   & $<$17.0               & N &                  &                         -2.28$^{2)}$  , -1.00$\pm$0.08 $^{3)}$ &                                                 &                       \\
 DM015        &               & III & Diskless   &                       & N &                  &                         -1.07$^{2)}$                           &                                                 &                       \\
 DM016        &               & III & Diskless   &                       & N &                  & -2.83 $\pm$0.10$^{1)}$, -5.45$^{2)}$                           &                                                 &                       \\
 DM017        &               & III & Diskless   &                       & N &                  &                         -0.25$^{2)}$                           &                                                 &                       \\
 DM018        &               & III & Diskless   &                       & N &                  &                         -2.64$^{2)}$                           &                                                 &                       \\
 DM019        &               & III & Diskless   & $<$17.0               & N &                  &                         -2.07$^{2)}$  , -2.44$\pm$0.23 $^{3)}$ &                                                 &                       \\
 DM021        &               & III & Diskless   &                       & N &                  &                         -1.73$^{2)}$                           &                                                 &                       \\
 DM022        &               & III & Diskless   &                       & N &                  &                         -2.37$^{2)}$                           &                                                 &                       \\
 DM023        &               & III & Diskless   &                       & N &                  &                         -2.04$^{2)}$                           &                                                 &                       \\
 DM024        &               & III & Diskless   &                       & N &                  &                          1.07$^{2)}$                           &                                                 &                       \\
 DM025        &               & III & Diskless   &                       & N &                  &                         -1.47$^{2)}$                           &                                                 &                       \\
 DM026        &               & III & Diskless   &                       & N &                  &                          0.39$^{2)}$                           &                                                 &                       \\
 DM027        &               & III & Diskless   &                       & N &                  &                         -3.64$^{2)}$                           &                                                 &                       \\
 DM028        &               & III & Diskless   &                       & N &                  &                          0.35$^{2)}$                           &                                                 &                       \\
 DM030        &               & III & Diskless   & $<$17.0               & N &                  &                         -1.0 $^{2)}$  , -1.89$\pm$0.15 $^{3)}$ &                                                 &                       \\
 DM031        &               & III & Diskless   &                       & N &                  &                          0.01$^{2)}$                           &                                                 &                       \\
 DM032        &               & III & Diskless   & $<$17.0               & N &                  &                         -5.5 $^{2)}$  , -4.98$\pm$0.49 $^{3)}$ &                                                 &                       \\
 DM034        &               & III & Diskless   &                       & N &                  &                         -2.64$^{2)}$                           &                                                 &                       \\
 DM035        &               & III & Diskless   &                       & N &                  &                         -1.67$^{2)}$                           &                                                 &                       \\
 DM037        &               & III & Diskless   &                       & N &                  &                         -0.08$^{2)}$                           &                                                 &                       \\
 DM040        &               & III & Diskless   &                       & N &                  &                         -3.63$^{2)}$                           &                                                 &                       \\
 DM042        &               & III & Diskless   &                       & N &                  &                         -2.94$^{2)}$                           &                                                 &                       \\
 DM043        &               & III & Diskless   &                       & N &                  &                         -2.60$^{2)}$                           &                                                 &                       \\
 DM045        &               & III & Diskless   &                       & N &                  &                          0.12$^{2)}$                           &                                                 &                       \\
 DM048        &               & II  & Thick      &                       & N &                  & -7.72 $\pm$0.19$^{1)}$, -8.61$^{2)}$                           &                                                 &                       \\
 ``           &               & II  & Thick      &                       & N &                  & -3.36 $\pm$0.11$^{1)}$, -8.61$^{2)}$                           &                                                 &                       \\
 DM052        &               & III & Diskless   &                       & N &                  &                          0.36$^{2)}$                           &                                                 &                       \\
 DM053        &               & III & Transition &                       & N &                  &                         -3.62$^{2)}$                           &                                                 &                       \\
 DM057        &               & III & Diskless   &                       & N &                  &                          0.15$^{2)}$                           &                                                 &                       \\
 DM061        &               & III & Diskless   &                       & N &                  &                         -4.23$^{2)}$                           &                                                 &                       \\
 DM062        &               & III & Diskless   &                       & N &                  &                         -4.05$^{2)}$                           &                                                 &                       \\
 DM063        &               & II  & Thick      &                       & N &                  &                         -4.76$^{2)}$                           &                                                 &                       \\
 DM064        &               & II  & Thin       &                       & N &                  &                         -2.52$^{2)}$                           &                                                 &                       \\
 DM065        &               & III & Diskless   &                       & N &                  &                         -2.67$^{2)}$                           &                                                 &                       \\
 DM066        &               & III & Diskless   &                       & N &                  &                         -1.0 $^{2)}$                           &                                                 &                       \\
 DM067        &               & III & Diskless   &                       & N &                  &                         -4.88$^{2)}$                           &                                                 &                       \\
 DM068        &               & III & Diskless   &                       & N &                  &                          0.04$^{2)}$                           &                                                 &                       \\
 DM069        &               & III & Diskless   &                       & N &                  &                         -0.49$^{2)}$                           &                                                 &                       \\
 DM070        &               & III & Diskless   &                       & N &                  &                         -0.90$^{2)}$                           &                                                 &                       \\
 DM071        &               & --  &            &                       &   &                  &                                                                &                                                 &                       \\
 C69-IRAC-001 &               & II  & Thick      &                       & N &                  & -4.79 $\pm$0.16$^{1)}$                                         &                                                 &                       \\
 C69-IRAC-002 &               & II  & Thick      &                       & N &                  & -9.05 $\pm$0.48$^{1)}$                                         &                                                 &                       \\
 C69-IRAC-003 &               & II  & Thin       &                       & N &                  & -6.31 $\pm$0.58$^{1)}$                                         &                                                 &                       \\
 C69-IRAC-004 &               & II  & Thick      &                       & N &                  & -11.65$\pm$0.25$^{1)}$                                         &                                                 &                       \\
 C69-IRAC-005 &               & II  & Thick      &                       & Y & -5.56$\pm$0.25   & -79.82$\pm$5.01$^{1)}$                                         & 780$\pm$25$^{1)}$                               &  -5.36;-5.60;-4.25 \\
 C69-IRAC-006 &               & II  & Thick      &                       & N &                  & -6.38 $\pm$0.13$^{1)}$                                         &                                                 &                       \\
 C69-IRAC-007 &               & II  & Thin       &                       & Y & -7.05$\pm$0.02   & -30.20$\pm$1.14$^{1)}$                                         & 561$\pm$84$^{1)}$                               &                       \\
 C69XE-009    &               & III & Transition?&                       & N &                  & -5.86 $\pm$0.28$^{1)}$                                         &                                                 &                       \\
 C69XE-040    &               & III & Diskless   &                       & N &                  & -4.91 $\pm$0.58$^{1)}$                                         &                                                 &                       \\
 C69XE-064    &               & III & Diskless   &                       & N &                  & -7.57 $\pm$0.34$^{1)}$                                         &                                                 &                       \\
 C69XE-072    &               & III & Diskless   &                       & N &                  & -4.64 $\pm$1.10$^{1)}$                                         &                                                 &                       \\
 C69XE-104c   &               & III & Diskless   &                       & N &                  & -8.63 $\pm$0.80$^{1)}$                                         &                                                 &                       \\
\end{longtable}

{\footnotesize
{\bf Notes:}\\
$^1$ Binary according to:\\
S08 - \citet{Sacco08}  \\
M08 - \citet{Maxted08} \\
$^2$ Infrared Class derived with the IRAC data in \citet{Barrado07} or \citet{MoralesPhD}.\\
$^3$ From \citet{Sacco08} or this work ($^{*)}$).\\
$^4$ According to H$\alpha$ EW and \cite{Barrado03} criterion after the analysis of the presence of disk.\\
$^5$ Measurement from:\\
\hspace{0.5cm} $^{1)}$; B11 - This work\\
\hspace{0.5cm} $^{2)}$; DM - \citet{DM99,DM01} \\
\hspace{0.5cm} $^{3)}$; S08 - \citet{Sacco08}  \\
}
\normalsize
\end{landscape}

}

\Online

\begin{appendix} 

\section{Particular sources}
\label{AP_PS}

In this appendix we provide further analysis for sources showing peculiarities in the properties studied in Sections~\ref{sec:rotvel} and~\ref{sec:acacc}. To keep consistency throughout the paper we have grouped the interesting sources from Section~\ref{sec:acacc} following the same subsection scheme.

\subsection{Rotational velocities}
\label{subsec:rotvel:ps}

{\bf LOri075:} This source has been classified as single-line spectroscopic binary (SB1) by \citet{Maxted08} (but no binarity signpost has been reported in \citealt{Sacco08}). According to \citet{Maxted08}, the spectral lines for this star show rotational broadening; they compared them to those of a narrow-lined star of similar spectral type and estimated a projected rotational velocity of v$\sin(i) \sim$ 65 km/s. They classified the source as SB1, but they also noted that there is an asymmetry in the cross-correlation function (CCF) in the form of a blue-wing, particularly when the measured radial velocity corresponds to a red-shift. Therefore they suggested that the fainter component in this binary was detected but unresolved in their spectra. 

\begin{figure}
\centering
\includegraphics[width=9.0cm]{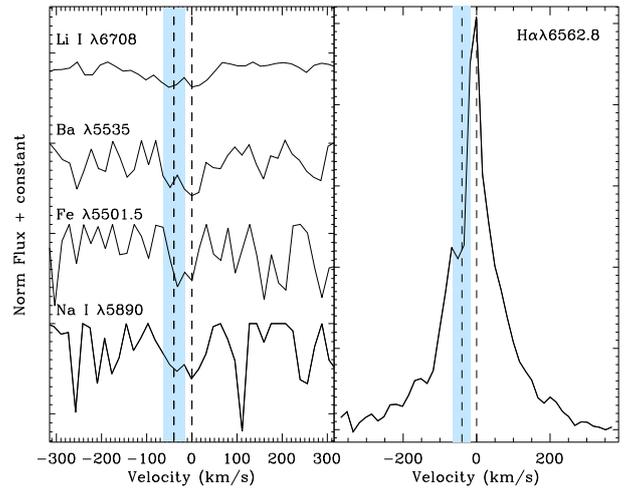}
\caption{{\bf Left:} Double peaked structure found in photospheric lines for LOri075. Rest frame velocity and mean relative velocity of the second peak are indicated with dashed lines. The shaded (blue) rectangle shows the $\pm 3 \sigma$ area of the second peak location. {\bf Right:} H$\alpha$ profile of the same source. Note how the secondary peak dashed line location (calculated from the photospheric lines) does not agree with the position of the peak (see text for details).}
\label{fig:LOri075SB}
\end{figure}

We have detected a double peaked structure in H$\alpha$ and Li I in our Magellan/MIKE spectra which made us believe that we had spectroscopically resolved the source. While the origin of double peak in Li I should be related to binarity, the H$\alpha$ one could arise from an accreting companion for example. Further research on the structure of some photospheric lines marginally confirms this idea. In Fig~\ref{fig:LOri075SB} we show the double peaked structure found in some of the most prominent photospheric lines for this object (given the low temperature of the source, these ``most prominent lines'' are still very weak). We have measured a mean relative velocity of $\sim$45 km/s ($\sigma \sim$ 9 km/s). We have synthesized a 3500 K (log(g) = 4.0) Kurucz spectra (an effective temperature $\sim$100 K hotter than the one estimated for our source, but the coolest temperature for the Kurucz collection) in the region of the Ba $\lambda$5535\AA~line with the same resolution and a rotational velocity close to the one derived by \citet{Maxted08} ($\sim$50 km/s). We have checked that the closest line in the synthetic spectra has a relative velocity of $\sim$110 km/s, much higher than those measured by us. We show on the right-hand side panel that the relative velocity derived for the photospheric lines does not agree with the one that would be measured from the H$\alpha$ profile. This fact and the weakness of the lines measured force us to consider the resolution of the binary as tentative. We must note anyway that the environmental H$\alpha$ component (see Fig~\ref{fig:HalphaFLAMES}) of the region or a possible accreting companion could change the relative velocity of the peaks of this emission line.

\subsection{Activity and accretion}

\subsubsection{Variability connected to activity}
\label{subsec:var:ps}

In our study of the H$\alpha$ variable sources, we find a sub-set of five objects for which the criterion from \cite{Barrado03} applied to spectra taken at different epochs provides contradictory results. While for some measurement of the object taken in one epoch (EW$(H\alpha)_1$), the source would be classified as accretor; for a different epoch measurement of the same object (EW$(H\alpha)_2$), the H$\alpha$ emission could be explained purely in terms of activity.

\begin{figure}
\centering
\includegraphics[width=9.0cm]{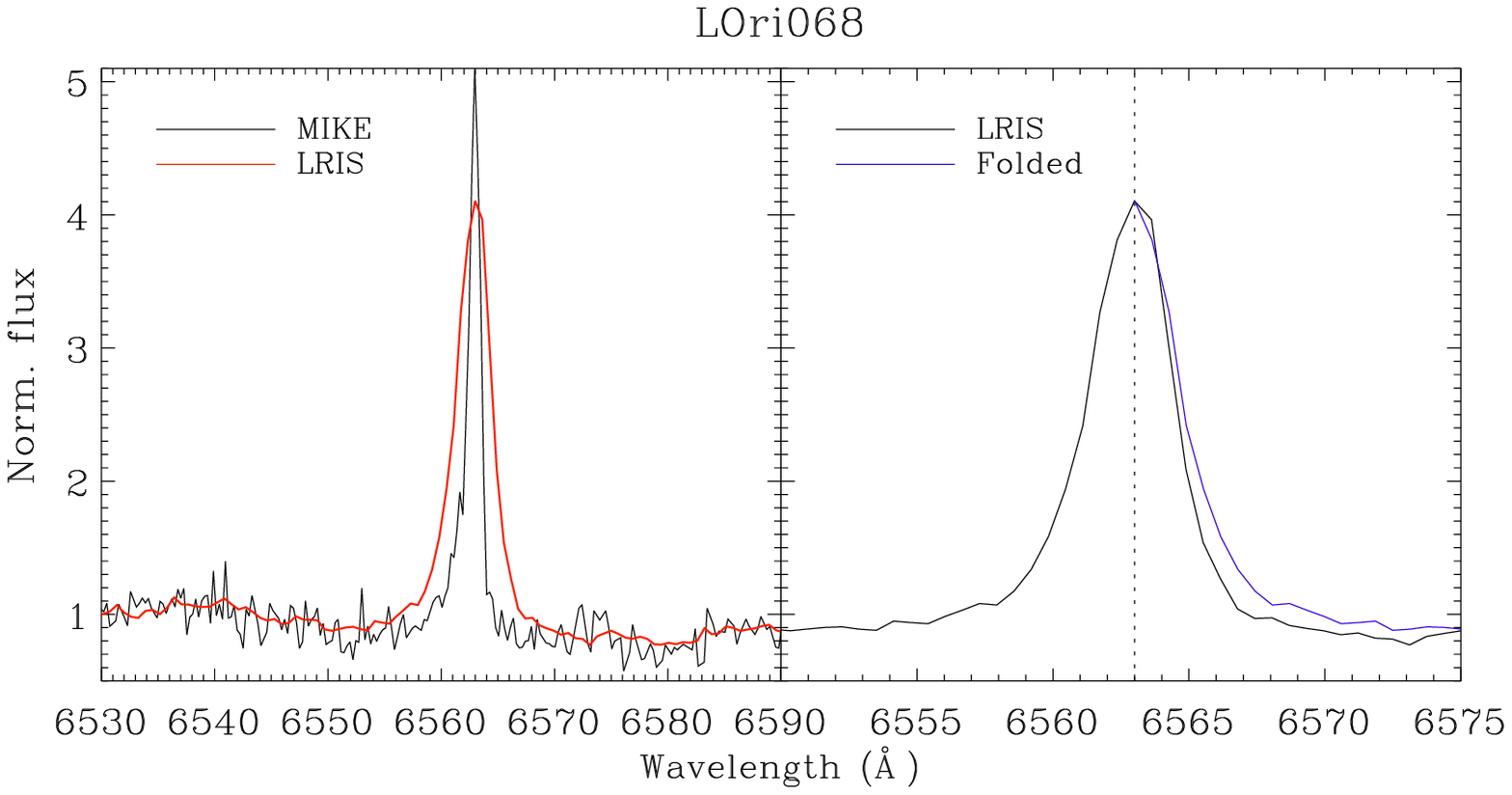}
\includegraphics[width=9.0cm]{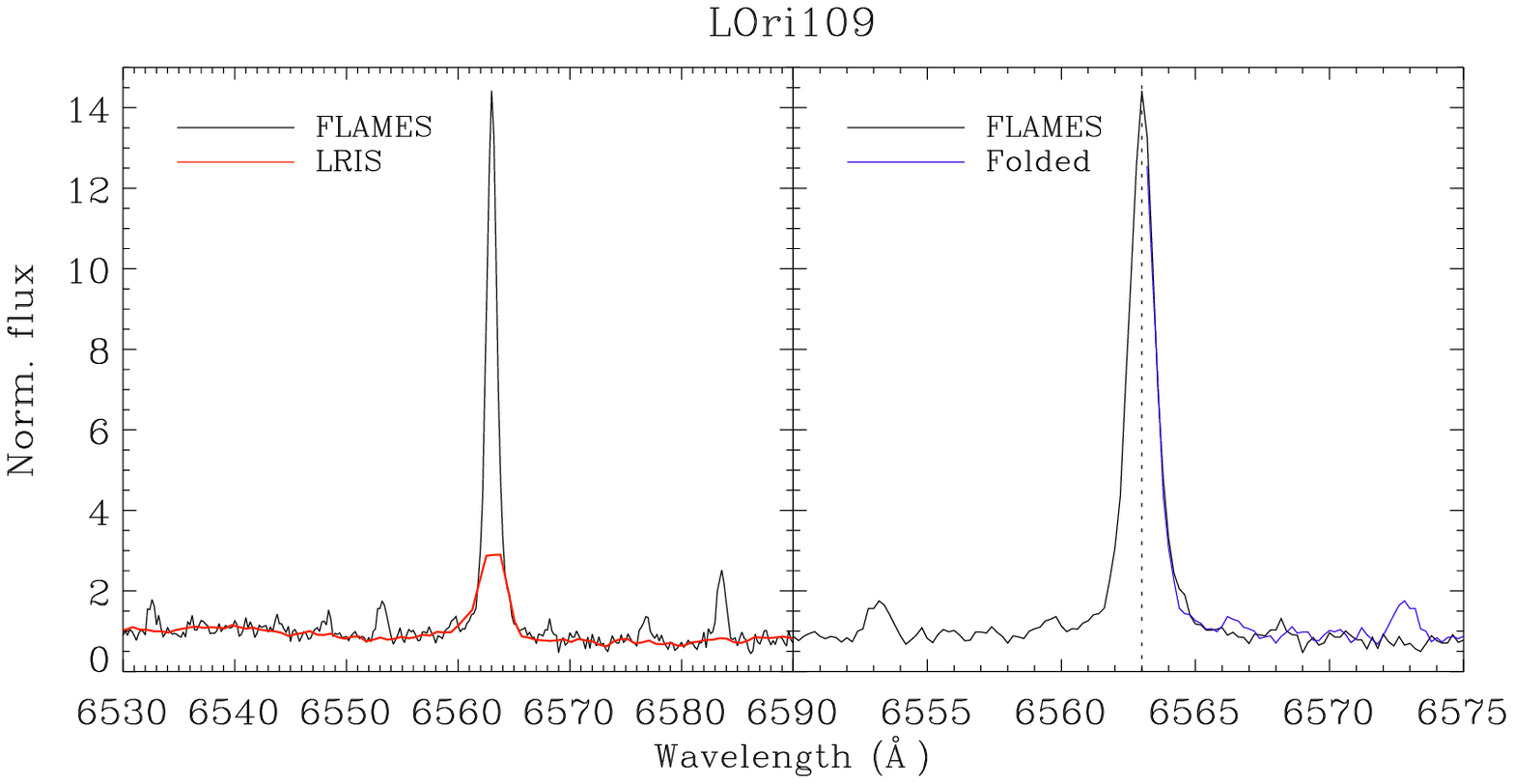}
\caption{Comparisons of the H$\alpha$ line profiles for different observations of the same two objects: {\bf Upper panel:} LOri068; one of the objects we suspect experimented a flare during the Keck/LRIS observations (R$\sim$2700). Some asymmetry can be seen in the line even though the resolution of the spectrum is moderate. {\bf Lower panel:} LOri109; another object suspected to have a flare for which no asymmetry in the line has been found.}
\label{fig:lineprofile}
\end{figure}

\begin{itemize}

\item {\bf LOri068} and {\bf LOri109} were observed twice during our campaigns and LOri068 was also observed by \cite{Sacco08}. Both objects are classified as diskless sources based on their IRAC slopes, and while our Li I measurements agree among themselves (and for the case of LOri068, with the one provided by \citealt{Sacco08}), in one spectrum for each source, the H$\alpha$ emission is much more intense than in the others. We believe those spectra were taken while the objects were experiencing a flare. In Fig~\ref{fig:lineprofile} we compare the line profiles of those objects in ``steady" and ``flared" states; and we show that, while in the case of LOri068, we can see some asymmetry in the line profile for the intense emission (which would indicate mass motion), that is not the case for LOri109 (even though for the latter the change in EW is much stronger).

\item {\bf LOri091} cannot be classified with certainty as  a variable source. There is one measurement of H$\alpha$ clearly off from the other two available, but that measurement corresponds to a TWIN spectrum of very poor S/N which translates into a large uncertainty on the continuum, and therefore a very large error-bar in the measurement.

\item {\bf LOri075} is an un-resolved (or marginally resolved, see subsection~\ref{subsec:rotvel:ps}) double system, and therefore variability in the measured H$\alpha$ is not surprising.

\item {\bf LOri080} is a puzzling case. We observed the object twice; in 2003 (at Las Campanas) and 2005 (at Calar Alto; see Paper I for a description of the instrumentation used in each case), and both measurements agree within the errors. These measurements place LOri080 at the border of being classified as accreting according to the saturation criterion (see next subsection), although the object shows no infrared excess in the IRAC data. No peculiarity has otherwise been found regarding the profiles of the lines in either spectra. On the other hand, \cite{Sacco08}, also observed LOri080 with FLAMES and found an EW for H$\alpha$ significantly lower.

\end{itemize}

\subsubsection{H$\alpha$ emission as a proxy for accretion}
\label{subsubsec:Haacc:ps}

In the process of determining accretion fractions, ratios and their relation with the disk properties, we encountered several particular cases that we describe below.

{\bf LOri161:} Is the brown dwarf from Fig~\ref{fig:HaSpT} that, even though its H$\alpha$ emission places it well above the saturation criterion, has not been classified as harboring a disk according to its SED. The issue with this very faint source is that it was not detected in IRAC channels three and four (5.8 and 8.0 micron, respectively). Since the sensitivity of these channels is lower than that of one and two (3.6 and 4.5 micron), it could be the case that this object indeed has a disk that we are not sensitive to and that is undergoing accretion. In that scenario, the estimated accretion rate according to the ${\rm FW}_{10\%}(H\alpha)$ would be $\sim 1\times 10^{-10}$ M$_{\odot}/yr$, which is much lower than the accretion rate derived for LOri156 ($\sim 9.5\times 10^{-9}$ M$_{\odot}/yr$), also a brown dwarf with the same spectral type and discussed later on in this section, but harboring an optically thick disk. 

An example of such a disk would be a transitional disk, where the excess would be only detectable at larger wavelengths. We checked the new release of the WISE catalog (in the preliminary version the source is not detected) and we found a counterpart within 1". Unfortunately, although the photometry at the largest wavelengths ($\sim$11 and 20 micron) shows a clear excess, these measurements have been classified as ``U" (upper limit), and therefore we cannot confirm that this source does harbor a disk.

{\bf C69XE-009:} An X-ray candidate from \cite{Barrado11} confirmed spectroscopically as C69 member in Paper I. This object is right at the limit of the saturation criterion; based on its SED it was classified as a candidate transition disk, but the linear fit to the mid-infrared slope is photospheric. Given that it is clearly an active object (detected in X-rays), and that the disk possibility is based on a very slight excess detected only in one infrared band, we assume that the H$\alpha$ emission has its origin in chromospheric activity and not accretion.

{\bf C69-IRAC-005:} Is the source from Fig.~\ref{fig:M_Macc} exhibiting the largest accretion rate based on the ${\rm FW}_{10\%}(H\alpha)$ ($\log(\dot{M}) = -5.56\pm0.25$) and the Spitzer/IRAC photometry suggests that it harbors an optically thick disk. This particular source was observed with CAFOS in low resolution mode, with a wider wavelength coverage than the other instruments used (see Paper I). Thus, we have been able to obtain a different estimation for the accretion rate based on the equivalent width measurement of the components of the CaII triplet (at 8498~\AA, 8542~\AA~ and 8662~\AA). This emission could be a sign of chromospheric activity too, as in the case of H$\alpha$; but the obtained equivalent widths for the triplet are too large; placing our measurements in the broad-line component of the unresolved line structures that is generally related to accretion (see \citealt{Comeron03, Mohanty05}). Furthermore, as in \citet{Comeron03}, the CaII triplet line ratios are very close to 1:1:1 (quite different from the 1:9:5 expected ratio for optically thin emission). 

We used the following equations to estimate the accretion rate from the CaII triplet (these equations were derived from the accretion line profile study by \citealt{Muzerolle98} and are further discussed in \citealt{Comeron03}):
\begin{equation}
\log(\dot{M}_{acc}) = -34.15 + 0.89 \log(F_{\rm CaII(\lambda8542)})
\end{equation}
\begin{equation}
F_{\rm CaII(\lambda8542)}=4.72 \times 10^{33} EW({\rm CaII(\lambda8542)}) \times 10^{-0.4(m_{\lambda}-0.54A_{V})}
\end{equation}
where $F_{\rm CaII(\lambda8542)}$ is the flux in the line, $m_{\lambda}$ is the magnitude of the star at $\lambda$8542, and $A_V$ is the visual extinction translated to the wavelength of the line of study using \citet{Fitzpatrick99} relations. Since the bluest photometric point that we had for this object is the 2MASS J magnitude, we used the best fitting model to the SED of the source as a scaling factor to estimate $m_{\lambda}$. On the other hand, according to the intrinsic colors by \citet{Leggett92} and our determination of the spectral type (M3), we find a very low $A_V$ value of 0.03 mag (quite lower than the average value of 0.36 mag derived for the cluster by \citealt{Duerr82}, but neither of them would significantly affect this estimation).

We obtained an accretion rate value of $\sim$3$\times$10$^{-7}$ M$_{\odot}$/yr; almost an order of magnitude lower than the one obtained based on H$\alpha$, which gives us an idea of the caveats of estimating accretion rates from measurements that can be well contaminated by activity or even by wind contributions.

On the other hand, even with the two estimations differing by such a large factor, this object still seems to be experiencing heavy accretion. We have compared its spectrum with that of C69-IRAC-002 (another M3 star, observed with the same setup, harboring a disk but with H$\alpha$ compatible with pure chromospheric activity and no CaII emission) looking for veiling emission, and no blue excess has been found in the source (further than a marginal excess right in the blue edge of the spectra that we think corresponds to an instrumental signature rather than a real excess). This result is not surprising since the wavelength coverage starts at 6200~\AA, and veiling in young stars is normally detected at bluer wavelengths. Therefore, we would need further spectroscopic measurements to confirm the presence of veiling in this source.

{\bf LOri050, LOri061 and LOri063} are the other sources with more than one estimation of $\dot{\rm M}_{\rm acc}$. \\
{\bf LOri050} is a spectroscopic binary according to \cite{Sacco08} and \cite{Maxted08}; it has been classified as Class II according to its mid-infrared photometry and we obtained two spectra with different instrumentation (see Paper I for details). According to the H$\alpha$ emission, in both cases, the object is above the saturation criterion. The estimated accretion rates for both measurements agree well within the errors ($\log(\dot{M}) =$ -11.09$\pm$0.05, -10.91$\pm$0.05). Therefore, we are observing a very interesting system with a total stellar mass of $\sim$0.3 M$_{\odot}$ and a circumbinary disk actively accreting.\\
For {\bf LOri063}, on the other hand, the two available accretion rate estimations (from \citealt{Sacco08} and this work) differ by more than an order of magnitude ($\log(\dot{M}) =$ -10.7$\pm$0.3, -11.87$\pm$0.07). LOri063 harbors an optically thick disk according to its IRAC photometry, and the change in the full width at 10\% of the flux in H$\alpha$ is also reflected in the change in EW of the line ($>$9\AA).\\
Finally, our measurement of {\bf LOri061} does not agree at all with that from \cite{Sacco08} (two orders of magnitude difference, $\log(\dot{M}) =$ -10.2$\pm$0.3, -7.65$\pm$0.05). We believe this difference arises from how sensitive the measurement of the ${\rm FW}_{10\%}(H\alpha)$ is to the local continuum determination. In Fig.~\ref{fig:LOri061} we illustrate the case graphically. While our automatic procedure (see Appendix A of Paper I for details) identifies a local continuum, the thick light green line, other routines fitting global continuum could base their measurements on the teal line. This difference in the determination of the ``real base" of the line yields the large discrepancy in the estimated accretion rate. We must note in any case that among our data on accretors, we do not have other sources where the H$\alpha$ profile can provoke this confusion in the continuum determination.

\begin{figure}
\centering
\includegraphics[width=7.5cm]{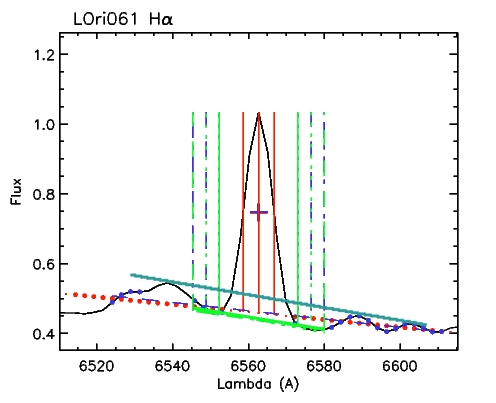}
\caption{Detail of the H$\alpha$ emission of LOri061 in the CAHA/TWIN spectrum. Note the dependence on the measurement of the full width at 10\% of the flux with the pseudo-continuum choice, in particular for the light green and the teal cases (for a complete description on the process to determine the different continuums see Appendix A of Paper I).
}
\label{fig:LOri061}
\end{figure}

{\bf LOri126, LOri140 and LOri156} are the three brown dwarfs (LOri126 is right at the limit between BD and very low mass star depending on the method used to estimate its mass, see Paper I) from Fig.~\ref{fig:M_Macc} exhibiting very large accretion rates ($\log(\dot{M}) = -8.78\pm0.10, -8.88\pm0.14, -8.02\pm0.10$, respectively). According to their mid-infrared slope, the three targets harbor optically thick disks. And according to their very large H$\alpha$ equivalent widths, they are well above the saturation criterion.

\begin{figure*}
\centering
\includegraphics[width=18.0cm]{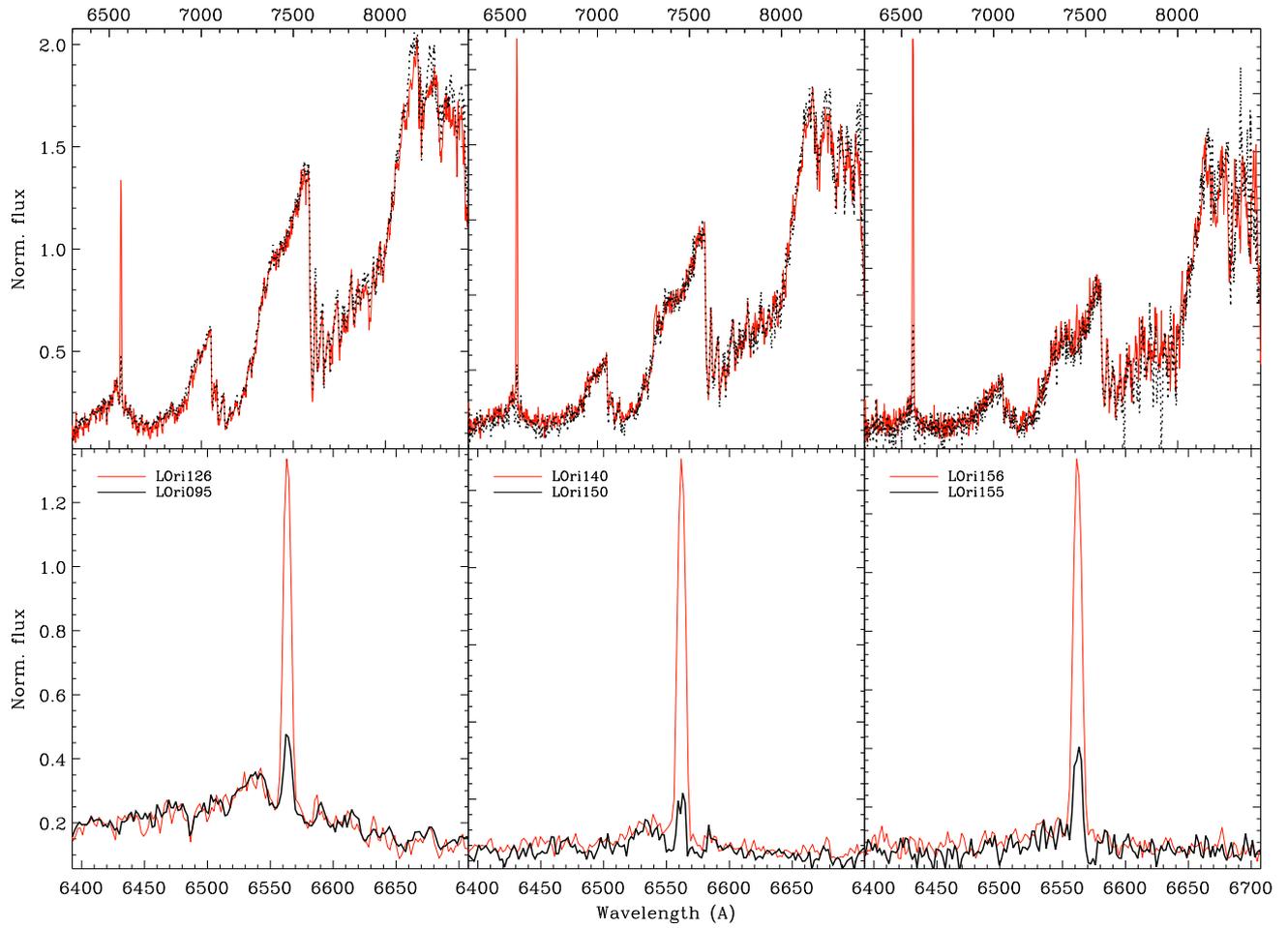}
\caption{{\bf Upper panel: }Comparison of accreting (red, very low mass / BD LOri126 and brown dwarfs LOri140 and LOri156) and non-accreting (black, with the same spectral type) members with low resolution spectra. Note the absence of veiling. {\bf Lower panel: } Detail of the H$\alpha$ emission o the same comparison.}
\label{fig:Accretors}
\end{figure*}

With such high accretion rates some veiling (due to excess emission from the accretion shock) could be expected in these sources (as it is the case for LS-RCrA 1, \citealt{Barrado04a}). In order to study this possibility, for each brown dwarf, we selected a non-accreting Class III source with very similar spectral type (one half subclass) and we compared the strength of several TiO molecular bands in both spectra. As can be seen in Fig~\ref{fig:Accretors} no significant differences are found in the continuum level of any pair of sources. In fact, in the three cases, $r_{\lambda}$, defined as $F(\lambda)_{\rm excess}/F(\lambda)_{\rm photosphere}$ is negligible. Whilst for LS-RCrA 1, \citet{Barrado04a} found that $r_{\lambda}$ varies from $\sim$1 to $\sim$0.25 for the wavelength range 6200--6750~\AA~we find a horizontal slope in this interval. The only cases where a linear horizontal $r_{\lambda}$ does not work are located on the very edges of the detector, and therefore we can conclude that no veiling is detected in any of the spectra in the studied wavelength range (this does not imply that some veiling cannot be present at bluer wavelengths).

\begin{figure}
\centering
\includegraphics[width=9.0cm]{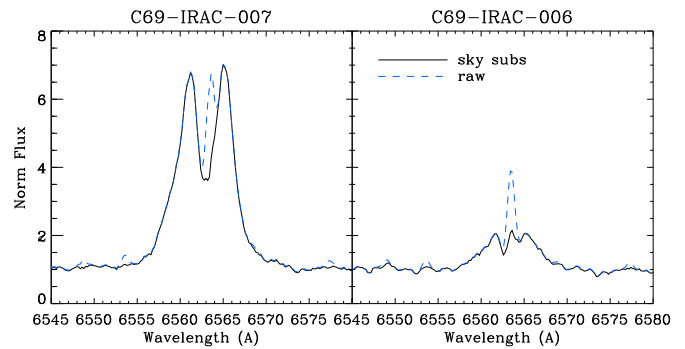}
\caption[Double peaked structure in the H$\alpha$ emission of C69WI1-9288 and C69WI1-2708.]{Double peaked structure in the H$\alpha$ emission of C69-IRAC-007 and C69-IRAC-006.}
\label{fig:doublepeak}
\end{figure}

{\bf C69-IRAC-006 and C69-IRAC-007:} Both sources are classified as Class II based on their IRAC photometry, and they show double-peaked structure of the H$\alpha$ emission as can be seen in Fig~\ref{fig:doublepeak}. Whilst the sky subtraction for C69-IRAC-007 worked very well, some residual could remain in the case of C69-IRAC-006 (although we do not see any structure on other, very narrow, ``sky lines''). 

This double peak is not present either in the other emission line detected in both spectra (He I) or in the absorption lines, discarding to a certain extent the possibility of these sources being spectroscopic binaries (SB2). We do not have an estimation of the rotational velocities of either given the resolution of the VLT/FLAMES observations (R$\sim$8000), but these almost symmetrical double peak structures in H$\alpha$ have been reproduced with models for higher mass stars with rapid rotation seen almost pole on (see \citealt{Muzerolle03} and references therein).

The accretion rate calculated for C69-IRAC-007 is shown in Table~\ref{tab:paramTOTAL} since this source fulfills the \citet{Barrado03} criterion ($\log(\dot{M}) = -7.05\pm0.02$). On the other hand, although the measured H$\alpha$ equivalent width of C69-IRAC-006 lies well below the saturation criterion of \cite{Barrado03}, the wide FW$_{10\%}$ measured ($\sim$190 km/s) places this object very close to the limit of accretors according to \citealt{Natta04}). In addition, note the resemblance of the H$\alpha$ profile of C69-IRAC-006 with that of Cha H$\alpha$2, an accreting brown dwarf, modeled in detail (and showing peculiarities attributed to the presence of an outflow) in \cite{Natta04}.

\end{appendix}

\end{document}